\documentclass{sig-alternate-10pt}
\usepackage{amssymb}
\usepackage{amsmath}
\usepackage{amsfonts}
\usepackage{graphicx}
\usepackage{bm}
\usepackage{tabularx}
\usepackage{url}
\usepackage[boxed]{algorithm2e}
\begin{document}

\title{Estimating Attendance From Cellular Network Data}

\numberofauthors{2}
\author{
\alignauthor
Marco Mamei\\
       \affaddr{Dipartimento di Scienze e Metodi dell'Ingegneria}\\
       \affaddr{University of Modena and Reggio Emilia, Italy}\\
       \email{marco.mamei@unimore.it}
\alignauthor
Massimo Colonna\\
       \affaddr{Engineering \& Tilab}\\
       \affaddr{Telecom Italia, Italy}\\
       \email{massimo.colonna @telecomitalia.it}
}


\maketitle

\begin{abstract}
We present a methodology to estimate the number of attendees to events happening in the city from cellular network data. In this work we used anonymized Call Detail Records (CDRs) comprising data on where and when users access the cellular network. Our approach is based on two key ideas: (1) we identify the network cells associated to the event location. (2) We verify the attendance of each user, as a measure of whether (s)he generates CDRs during the event, but not during other times. We evaluate our approach to estimate the number of attendees to a number of events ranging from football matches in stadiums to concerts and festivals in open squares. Comparing our results with the best groundtruth data available, our estimates provide a median error of less than 15\% of the actual number of attendees.
\end{abstract}



\section{Introduction}

The widespread diffusion of mobile phones and cell networks provides
a practical way to collect geo-located information from a large user
population. The analysis of such collected data is a fundamental
asset in the development of pervasive and mobile computing
applications, including location-based services, traffic management,
urban planning, and disaster response \cite{Cal11,Fer11,Fer14,Bec13,Zam12,Leo14,Lat13}.

In this work, we explore the use of anonymized Call Detail Records
(CDRs) from a cellular network to estimate the number of attendees
to large events happening in the city.

Each CDR contains information such as the time a mobile phone accesses the network (e.g., to send/receive calls and text messages), as well as the
identity of the cell tower with which the phone was associated at
that time. CDRs can serve as sporadic samples of the
approximate locations of the phone's owner.

On the basis of such location samples, we try to understand if a
user was attending a given event and estimate the number of
attendees on that basis.

While in some contexts, the number of participants can be deducted
also by other means (e.g., ticketing information), there are many
scenarios in which counting the attendance is problematic (e.g.,
events held in open squares, parades, flash-mobs) and an estimate on
the basis of cellular network data is highly valuable.

Estimating events' attendance has a number of practical and useful
applications.

On the one hand, it is an important information for the local
government and organizers in that it is at the basis of event's
planning and resource prioritization. In addition, since CDRs
allow to track the movements of individual users, it is possible to
understand where attendees come from and where they go after the
event. This naturally supports traffic and road management.

On the other hand, such kind of information, can support
advertisement systems \cite{Que11} by providing accurate audience
measurements. Also in this case, the possibility of tracking users
would open to advanced applications for the provisioning of highly
personalized advertising and marketing schemas. Despite users' hashed ids do not allow to identify the real person behind a phone, this opens a number of privacy concerns. While some research addressing such concerns exist \cite{Mir13,Bas14}, we will not tackle privacy problems in this paper focusing on the attendance estimation problem only.

While a number of existing works deal with the problem of
discovering and analyzing events on the basis of cellular network
data (see Related Work section), the problem of actually estimating
the number of attendees is largely unexplored. In particular, to the
best of our knowledge, there are not published results of the
accuracy of attendance estimation using CDRs.

The goal of this paper is to present such an estimation procedure.
In particular, in Section 2 we present a naive approach to estimate
the attendance and illustrate why it does not work properly. In
Section 3 we present our methodology. In Section 4, we evaluate our
approach to estimate the number of attendees to football matches in
stadiums, in which reliable groundtruth data were available.
Section 5 discusses how to improve performance on the basis of the knowledge of multiple events in the area.
Section 6 presents related work. Eventually, Section 7
concludes and discuss some future avenues for improvement.

\section{Naive Approach}

Before illustrating the proposed methodology, we want to show the
main problem that complicates the task of estimating the number of
attendees.

A naive approach to address such an issue would be to just count the
number of users who generate CDRs in cells covering the
event's location area during the event time. In particular, we tried to
apply the naive approach to estimate the number of attendees to
football matches in two different stadiums in Turin, Italy. We
defined the area associated to each stadium as a circle centered in
the stadium with a fixed radius of 100m. Then, we record all the CDRs
produced in the network cells that overlap with the stadiums' area
at the event time. We then counted the number of individual users.

\begin{figure}[t]
\begin{center}
\includegraphics[width=1.0\columnwidth]{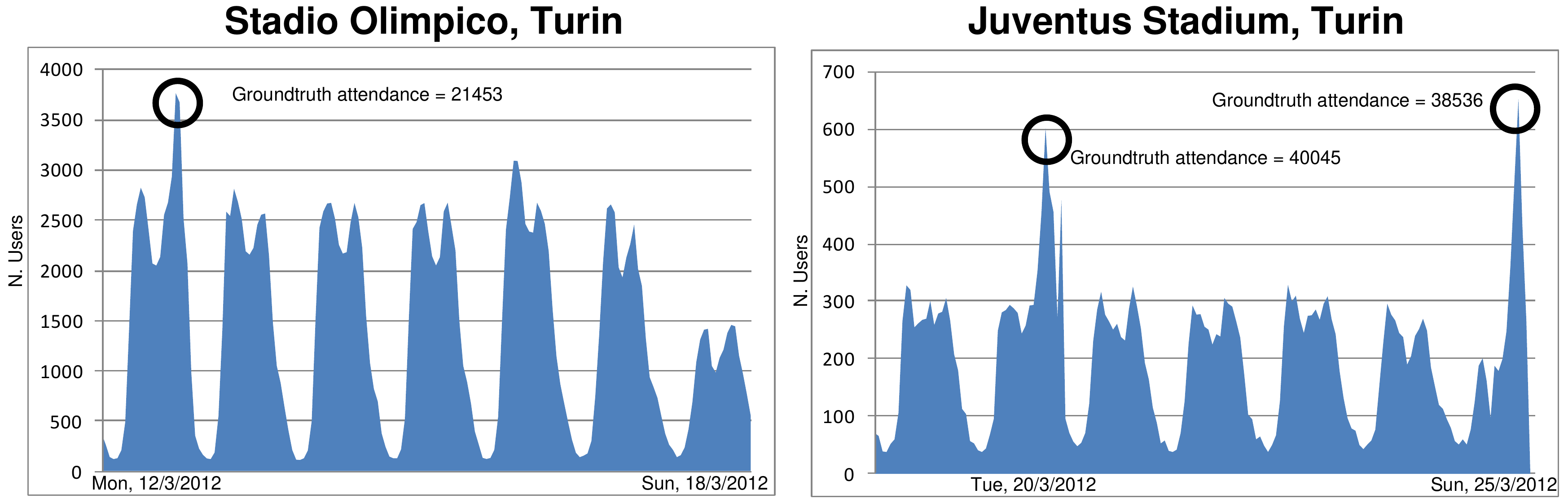}
\end{center}
\caption{Hourly count of users generating CDRs in the area
associated with the stadiums (Stadio Olimpico on the left, Juventus
Stadium on the right). The problem is in the negative correlation
between groundtruth and CDR counts: large events appear to be
smaller than ``small'' ones.} \label{fig:naive}
\end{figure}

Figure \ref{fig:naive} illustrates the result. The graphs represent
the hourly count of users in the area associated with the stadiums
(Stadio Olimpico on the left, Juventus Stadium on the right). We
also highlighted football matches taking place in the stadiums with
also groundtruth estimates for the number of attendees.

It is rather easy to see that the naive approach is highly
ineffective. For example, the match that happened on March 12, 2012
at the Stadio Olimpico is reported to have 21453 attendees and a CDR
users' count with a peak of about 3700. In contrast, the match that
happened on March 20, 2012 at the Juventus Stadium is reported to
have a double number of attendees (40045), while a CDR users' peak
of about one-sixth (600).

The problem with these numbers is not in the discrepancy between
groundtruth and CDR counts. This can be naturally explained by the
fact that not all the users use the phone during the match, and
by the fact that not all of them adopts the same carrier providing
the data for this analysis.

The problem is in the negative correlation between groundtruth and
CDR counts: large events (happening at the Juventus Stadium) appear
to be smaller than ``small'' ones (happening at the Stadio
Olimpico).

The reason for such a negative correlation can be easily found in
the geography of the city. Stadio Olimpico is right in the city
center. Juventus Stadium is in the suburbs. Accordingly, while
network cells around Juventus Stadium are likely to measure CDRs
coming from the stadium itself, network cells around Stadio Olimpico
overlap with a number of other relevant places and businesses in the
city center thus inflating the result.

\begin{figure}[t]
\begin{center}
\includegraphics[width=0.8\columnwidth]{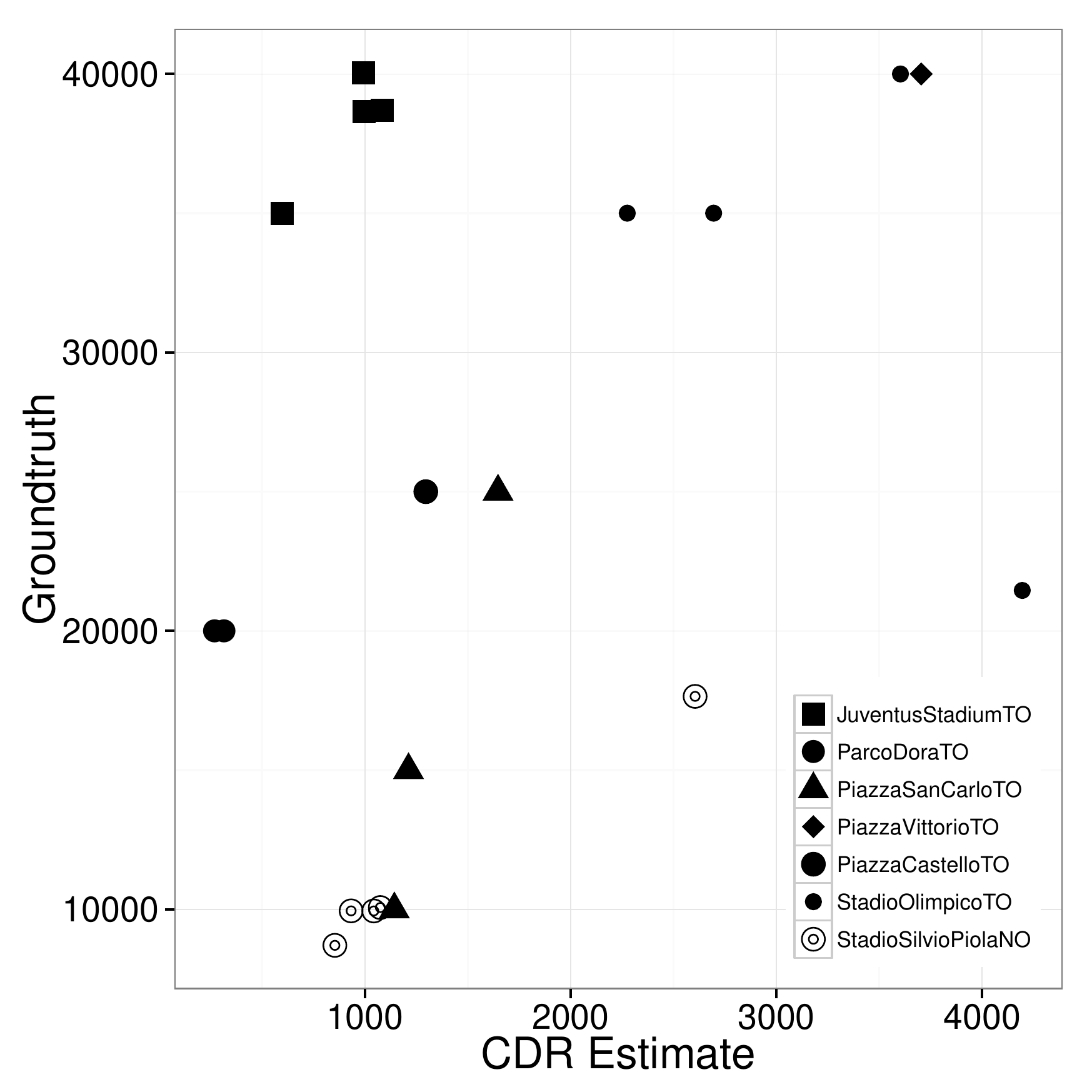}
\end{center}
\caption{Correlation result using the naive approach. It is easy to see that there is almost no correlation ($r^2 = 0.016$) among CDR count and groundtruth.} \label{fig:naive-correlation}
\end{figure}

More in general, Figure \ref{fig:naive-correlation} shows correlation results -- using the naive approach --  for a number of events covered by our dataset. Each point represents an event: the x-coordinate is the CDR estimate for attendance, while the y-coordinate is the groundtruth attendance. It is easy to see that there is almost no correlation ($r^2 = 0.016$) between the two estimates, so the naive approach is highly ineffective. Our goal is to identify a mechanism to create a strong {\it
positive} correlation between groundtruth and CDR counts. Once this
result is achieved, a simple linear regression can scale up CDR
counts to the actual attendees estimate.

\section{Methodology}

\begin{figure}[t]
\begin{center}
\includegraphics[width=0.99\columnwidth]{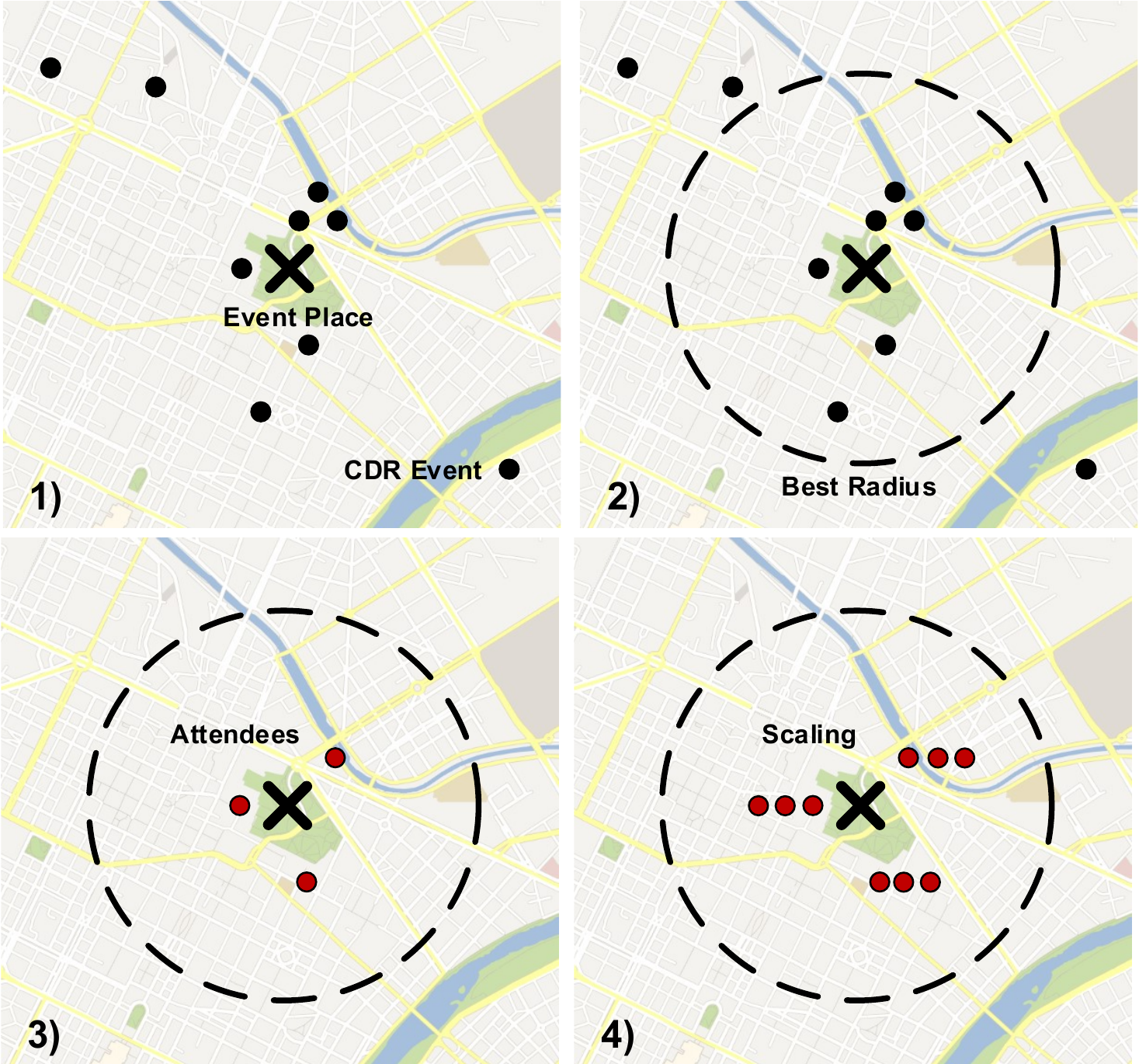}
\end{center}
\caption{Proposed methodology to estimate event's attendance. {\bf
1)} We collect CDRs generated around the event area is selected. {\bf 2)}
We compute the radius best describing the event area.
{\bf 3)} The number of users who generate CDRs at the event time,
but who do not (usually) generate CDRs at other times is recorded.
{\bf 4)} This number is then scaled according to a linear regression to find
the actual attendance estimate.} \label{fig:methodology}
\end{figure}

To overcome the above limitations, we developed a specific
methodology to deal with attendance estimation (see Figure
\ref{fig:methodology}). In particular: {\bf (1)} We collect all the
CDRs generated around the event area. {\bf (2)} We identify the
radius within which are all the cells whose traffic can
be associated to the area where the event takes place. {\bf (3)} On
the basis of the identified cells, we count the number of users who
generate CDRs at the event time, but who do not (usually) generate
CDRs at other times. Finally, on the basis of such data from a
number of events, we set up a linear regression to estimate the
number of attendees.

In the following subsections we describe in detail the above steps.

\subsection{CDR Data}

We obtained a large set of mobility data from an Italian telecom operator. In particular, we analysed data from two regions of Italy (Piemonte and Lombardia inhabited by about 15 millions people), spanning 16 months (March 2012 -- June 2013) during which we analysed several events ranging from football matches in stadiums to concerts and festivals in open squares.

Mobility data is obtained from Call Detail Records (CDRs) and Mobility Management (MM) procedure messages (i.e., IMSI attach/detach and Location Update) \cite{Rah93}. CDRs are routinely collected by cellular network providers for billing purposes. A CDR is generated every time a phone places or receives voice call or a text message. The IMSI attach/detach procedure marks the phone as attached/detached to the network on power up/power down of the phone or SIM inserted/removed. Location updates are messages exchanged for keeping the network informed of where the phone is roaming. CDR and MM messages are read on network interfaces through specific probes and also contain the identity of the phone, the identity of the cell through which the phone is communicating and the related timestamp. As MM messages, for the purposes of our study, contain the same information as CDRs, for simplicity of writing we will refer to all these data as CDRs.

In the context of this work, all this information serves as sporadic samples of the approximate locations of the phone's owner. Specifically, the user's location is given in terms of the cell network antenna the user was connected with. The area covered by a given antenna sector can be approximated by a circle with a given center and radius. In Figure \ref{fig:cdr-table} it is shown the structure of a CDR. Each record comprises a user (hashed) id , the MCC (Mobile Country Code) representing the country where the SIM card has been registered, the timestamp of the CDR, the code of the cell tower and the coordinates and coverage radius of the cell tower. Thus, the spatial resolution of CDR localization is the cell radius. Similarly to \cite{Cac12}, in our work we take into considerations different sectors for different antennas. Each sector is refereed to as an individual cell and approximated with a circle.

It is worth noticing that differently from a number of other works we do not estimate the coverage of a cell network by using Voronoi tessellation. We stick to the simpler representation of a cell being represented by a circle with a given center and radius. In \cite{Ulm13}, it is shown that the approach do not change the user's location accuracy.

\begin{figure}
\small
\begin{tabular}{ | c | c | c | c | c | c | c |}
\hline
{\bf User} & {\bf MMC} & {\bf Time} & {\bf Cell} & {\bf Coord} & {\bf Radius} \\ \hline
3dd2b & 222 & 7346286 & 123 & (41.2,13.9) & 450 \\
\hline
\end{tabular}
\caption{Structure of our CDR dataset. Every time a user send or receive calls and text messages we generate one CDR with information about the user (hashed) id, the MMC (Mobile Country Code), the timestamp of the CDR, the code of the cell tower and the coordinates and coverage radius of the cell tower.} \label{fig:cdr-table}
\end{figure}

\begin{figure}
\begin{center}
\includegraphics[width=0.48\columnwidth]{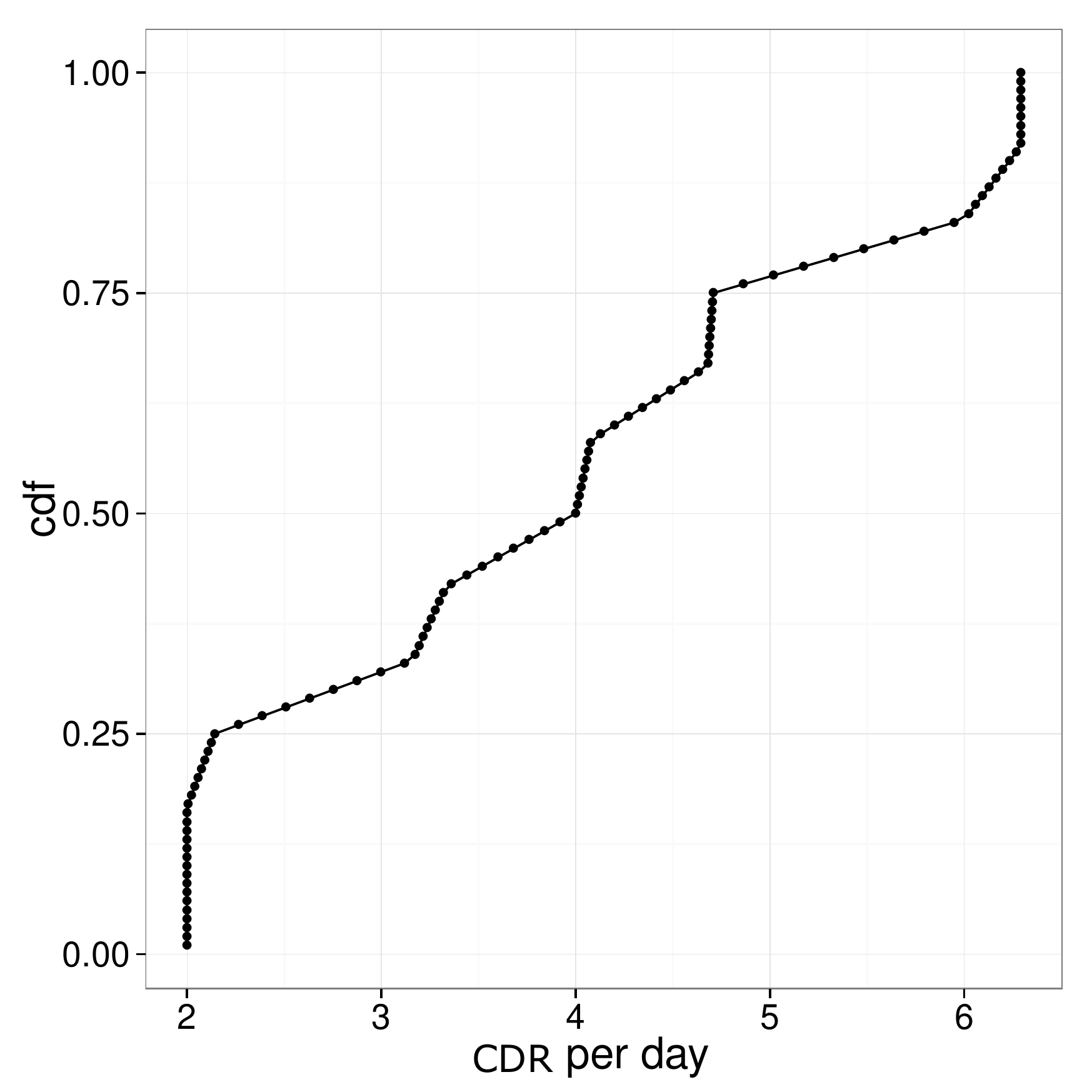}
\includegraphics[width=0.48\columnwidth]{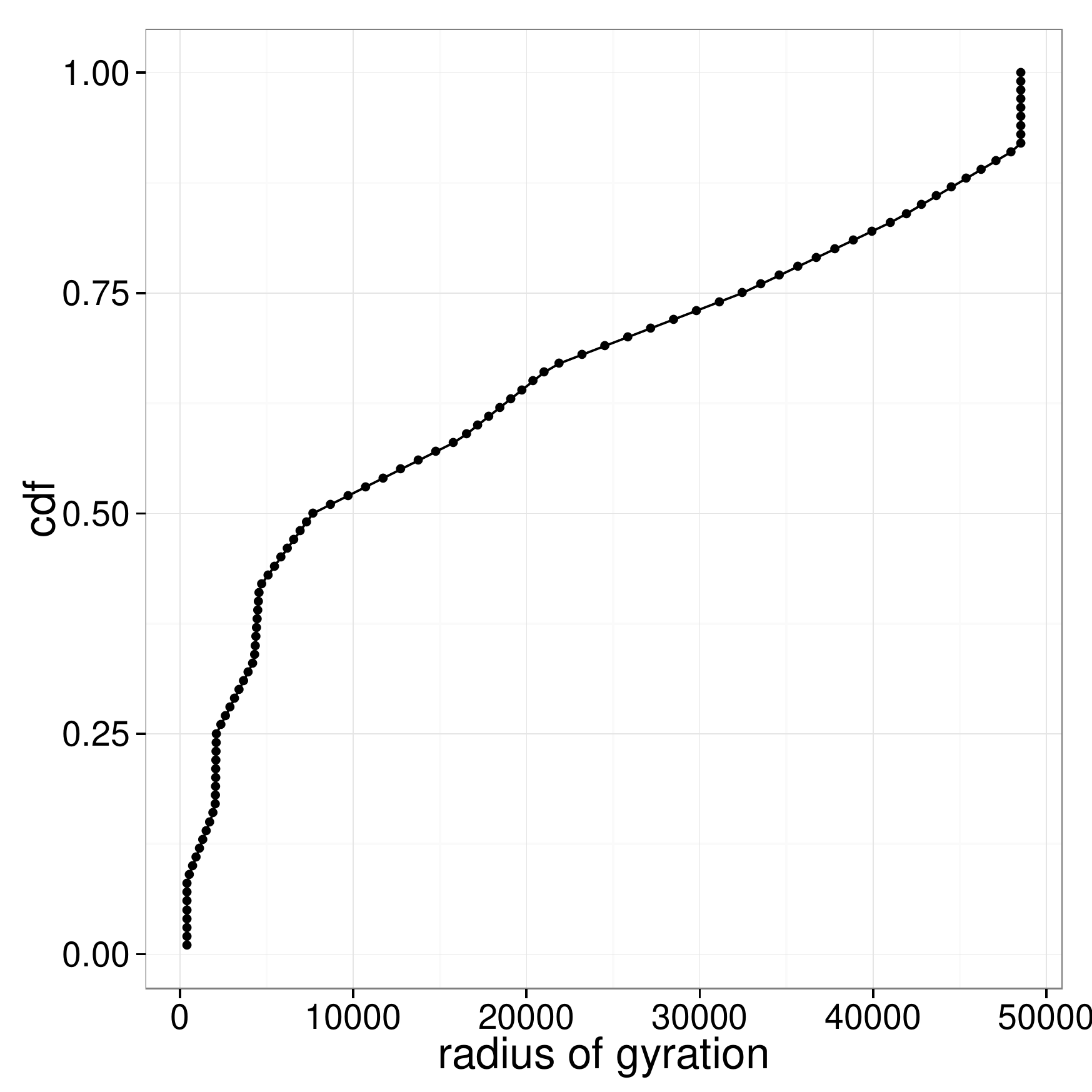}
\end{center}
\caption{(left) Daily average number of CDRs produced for a given percentile of users. (right) Radius of gyration for a given percentile of users.} \label{fig:cdr-stats}
\end{figure}

Figure \ref{fig:cdr-stats} illustrates some key statistics of our data.
Figure \ref{fig:cdr-stats}-left illustrates the daily average number of CDRs produced for a given percentile of users. While the average number of CDRs per day is rather limited, we monitor a large user population comprising more than 4 million persons. In addition, as discussed in Section 3.3, CDRs are not evenly spread across all the days and across the 24 hours. So, we actually have more location samples in the time frame where events actually happen.

Figure \ref{fig:cdr-stats}-right illustrates the radius of gyration for a given percentile of users.
The radius of gyration is a synthetic parameter describing the spatial extent of user traces. It is defined as the deviation of user positions from the corresponding centroid. It is given by:
$r_g = \sqrt{\frac{1}{n}\sum_{i=1}^n(p_i - p_{centroid})^2}$ where $p_i$ represents the $i^{th}$ position recorded for the user and $p_{centriod}$ is the center of mass of the user's recorded displacements obtained by: $p_{centroid} = \frac{1}{n}\sum_{i=1}^n(p_i)$. It is possible to see that almost half of the user are urban dweller with $r_g$ less than 10Km. Users in the  ($50^{th}$-$75^{th}$) percentiles can be associated to urban commuters as the diameter of peri-urban areas of main cities in the region is about 25-30Km. Users beyond the $75^{th}$ percentile are associated to commuters travelling region-wide.

\begin{figure}
\begin{center}
\includegraphics[width=0.65\columnwidth]{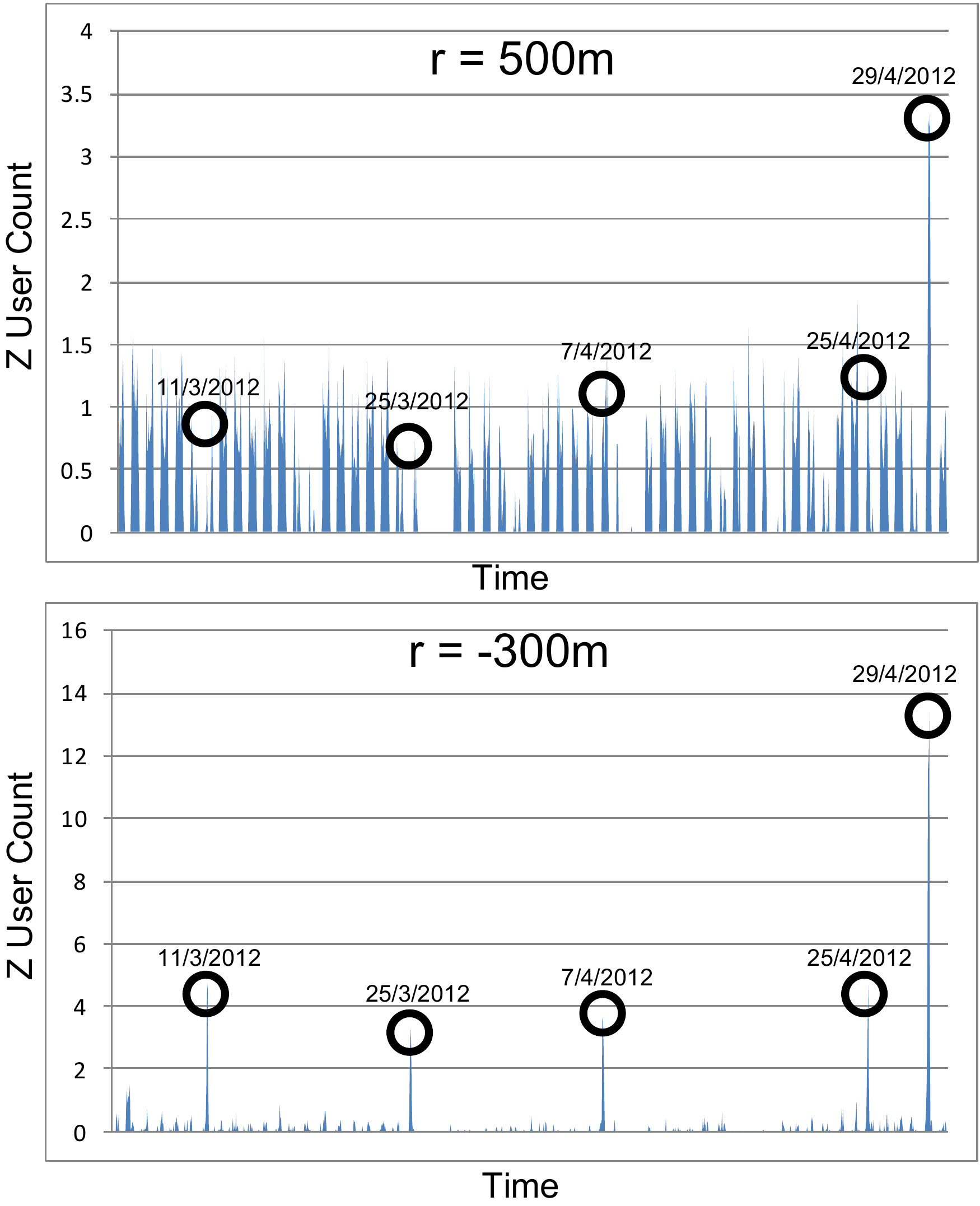}
\end{center}
\caption{Identification of the best radius to model the event area.
If the radius is too large ({\bf top}) the events' structure cannot
be identified properly. With a proper value of the radius ({\bf
bottom}) outlier in the CDR counts correspond to the events.}
\label{fig:bestr}
\end{figure}

\subsection{Best Radius}
\label{sec:radius}

As discussed in Section 2, determining the cells that are relevant
for the events generated in a given area is a fundamental task.
Otherwise it is possible that the cells being considered will
include CDRs actually produced elsewhere, or will miss CDRs that were actually produced in the proper area.

To tackle this problem, we model the event area as a circle with
center $c$ - where the event takes place, and with radius $r$. A
cell with center $b$ and radius $rc$ is considered relevant for
the event if: $dist(c,b) < r+rc$. Where $dist$ is the geographic
distance between the points. In other words, a cell is relevant if
it overlaps with the circle representing the event area.

The problem of determining the relevant cells is thus shifted to the
problem of identifying a proper radius $r$ for the event area. It is
important to notice that we could also select $r < 0$ to impose the
fact that a cell has to overlap to the center of the event by a
certain amount to be considered as relevant.

To solve this issue, our approach starts from the basic
consideration that the plot of the number of CDRs generated from the
event area should have a spike (i.e., an outlier) when the event takes
place, as the events -- we are interested in -- will typically
attract a large number of people.

For example, Figure \ref{fig:bestr} illustrates the {\it z-score}
for the hourly count of users producing CDRs around a stadium (Stadio Silvio Piola, Novara, Italy).
In the top graph, the stadium area is modeled as a circle with radius $r = 500 m$.
In the bottom graph, the stadium area is modeled as a circle with radius $r = -300 m$ (see above discussion on negative radii). It is easy to see that adopting $r=500m$ fails to capture the events'
structure in that events are not clear outliers. On the contrary with $r=-300m$ it is possible to precisely identify events (i.e., all the events have values larger than 3).

In this context, $r=-300m$ would be a suitable radius to describe the event area. This is probably due to the fact that the stadium is close to
other relevant places and businesses. Taking large values of $r$ bias the CDR count by considering also CDRs generated in these other places. Instead, a low value of $r$ selects only relevant CDRs. It is also possible to see that the outlier associated to the event on 29/4/2012 is readily visible even with $r = 500 m$. The football match that happened on that date, in fact, attracted almost the double of people (17650 persons vs. stadium's average of 9370). Such an event would be better represented by a larger radius (the more the people, the more the cells nearby the stadium get saturated and rely the network connection to farther cells).

On the basis of the above considerations, we developed an approach to identify the best radius describing the event area.
For each event happening at a location  with center $ec$ starting at time $st$ and ending at $et$, we propose the the following approach. For the sake of clarity, we present the approach in two different steps.\\

\noindent
{\bf STEP 1.}

\begin{enumerate}

\item  For different values of $r$ in $r_{min}$, $r_{max}$, we extract the CDRs in the
event area ($cdr[]$).

\item For each $r_k$, we compute the hourly count of users who
generate CDRs in the area during the event time. We call $x_k$ such a count.

\item We then compute the {\it z-score} of the $x_k$ values in the event time frame. More in
detail, we computed the hourly count of users who generate CDRs in the area during the event time, but in $i$ days before the event (we considered 6 days before).
We then computed the mean $\mu_k$ and standard deviation $\sigma_k$ of this count. On this basis we computed the z-score $z_k = (x_k - \mu_k)/\sigma_k$.
The result is a series of values $z_k$ measuring how extreme the CDR count were during the event (considering given radius $r_k$).

\end{enumerate}

\begin{algorithm}[h]
 \KwData{$cdr[\;]$, $ec$, $st$, $et$}
 \KwResult{$z[\;]$}
 \ForAll{$r_k \in [r_{min}, r_{max}]$ }{
    $x_k = countUsers(cdr[\;],ec,r_k,st,et)$

    \ForAll{i $\in$ [0, 6]}{
        $y_{ik} = countUsers(cdr[\;],ec,r_k,st - i \cdot days, et - i \cdot days)$
    }

    $\mu_k = mean_i(y_{ik})$
    $\sigma_k = sdt.dev_i(y_{ik})$

    $z_k = (x_k - \mu_k)/\sigma_k$
 }
 \caption{Radius Extraction - Step 1}
 \label{algo:radius-step1}
\end{algorithm}

Algorithm \ref{algo:radius-step1} presents a more formal description of the approach.

\begin{figure}
\begin{center}
\includegraphics[width=0.48\columnwidth]{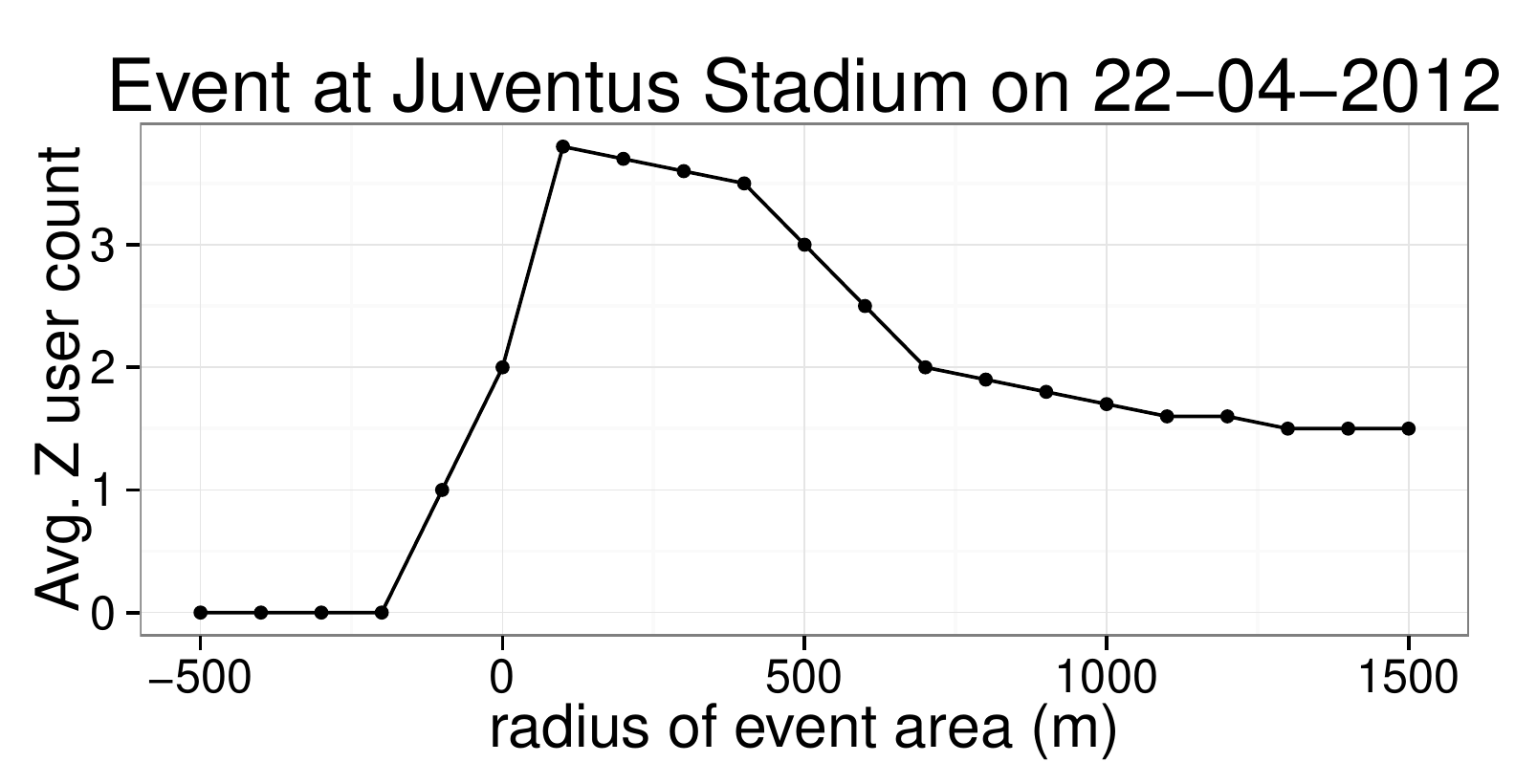}
\includegraphics[width=0.48\columnwidth]{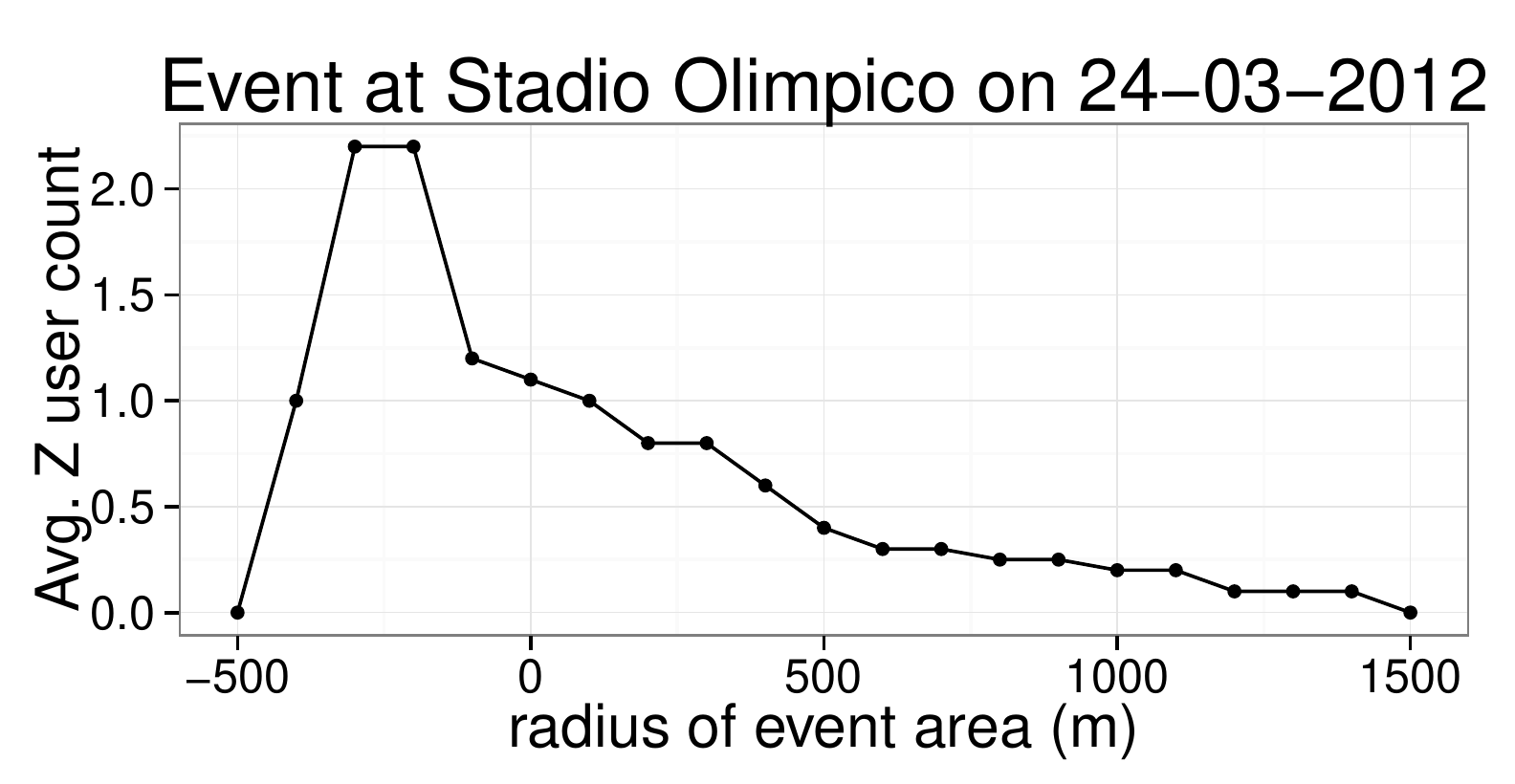}
\end{center}
\caption{Graph showing for each $r_k$ how much the area had an unusually high number of people during the event.}
\label{fig:zXradius}
\end{figure}

The result is a graph showing for each $r_k$ how much the area had an unusually high number of people during the event. Figure \ref{fig:zXradius} shows the result for two events.
It is possible to see that once the area is properly identified, the z-score clearly identifies that something unusual is taking place there ($z_k = 3.7$ with a radius of about 300m for the event on the left, $z_k = 2.2$ with a radius of about -200m for the event on the right).\\

\noindent
{\bf STEP 2.}
On the basis of the graph showing the average z-score for different radii, we have to identify the actual {\it best} radius.
Contrarily to a naive approach, selecting the radius associated to the maximum $z$ is not an effective option. This approach would be strongly biased to small radii that always exhibit large z-scores. In fact, even if the event area is large, any (smaller) area contained in there would have a z-score that is likely to be higher than the whole evet area, as it comprises only those cells that are really in the middle of the event. Accordingly, we adopted the following solution. See also Algorithm \ref{algo:radius-step2}.

\begin{enumerate}

\item For each $r_k$, we normalize the $z_k$ values by a factor representing the event area. The idea is that a large $z$ over a small area around the event's location should be favoured with respect to a a large area possibly comprising also other events. In particular, we divide each $z_k$ by the sum of the radii of the network cells associated with the $r_k$ area. More formally, calling $nc_k$ the set of the network cells within the event area defined by $r_k$, and calling $nc.r_i$ their radii, the our normalized z value $\hat{z}$ is computed as: $\hat{z}_k = z_k / \sum_{i \in NC_k} R_i$

\item Finally, we compute the best radius as the average of the $r_k$ values weighted by the associated normalized z-scores.

\end{enumerate}

\begin{algorithm}[h]
 \KwData{$z[\;]$}
 \KwResult{$bestR$}
 \ForAll{$r_k \in [r_{min}, r_{max}]$ }{

    $\hat{z_k} = z_k / \sum_{i \in nc_k} nc.r_i$

    $bestR = \frac{\sum_k r_k \cdot \hat{z}_k}{\sum_k \hat{z}_k}$
 }
 \caption{Radius Extraction - Step 2}
 \label{algo:radius-step2}
\end{algorithm}

\subsection{Attendance Estimator}

Once the event area has been identified, we need a mechanism to
count precisely the number of users who attended the event. Since we
do not know what the user was doing in the event area, we estimate
the probability of the user presence as proportional to the fraction
of time in which the user was there during the event, and inversely
proportional to the fraction of time in which the user was there
outside of the event time \cite{Tra11}. This latter point is important to
eliminate users that live or work in the event area and so are in
there independently of the event.

As a first step, we tried to characterize the individual calling
activity and verified that it is frequent enough to allow monitoring
the users' location with a fine enough resolution. For each user, we
measured the inter-CDR time - i.e., the time interval between two
consecutive network connections (similar to what has been done in
\cite{Gon08,Cal11}). Focusing on a given event (e.g., a football game held at the Juventus
Stadium in Turin on March 20 2012), we performed some measures. The
average inter-CDR time measured for the population of possible
attendees (users who generate at least one CDR in the event area
during the event time) was 241 minutes. This number is large because
it considers the whole daily lives of that users, thus also spanning
night gaps. We also measured the average inter-CDR time
considering only CDRs generated during at the event time. With that
assumption the average inter-CDR time reduces to 52 minutes.

\begin{figure}
\begin{center}
\includegraphics[width=0.8\columnwidth]{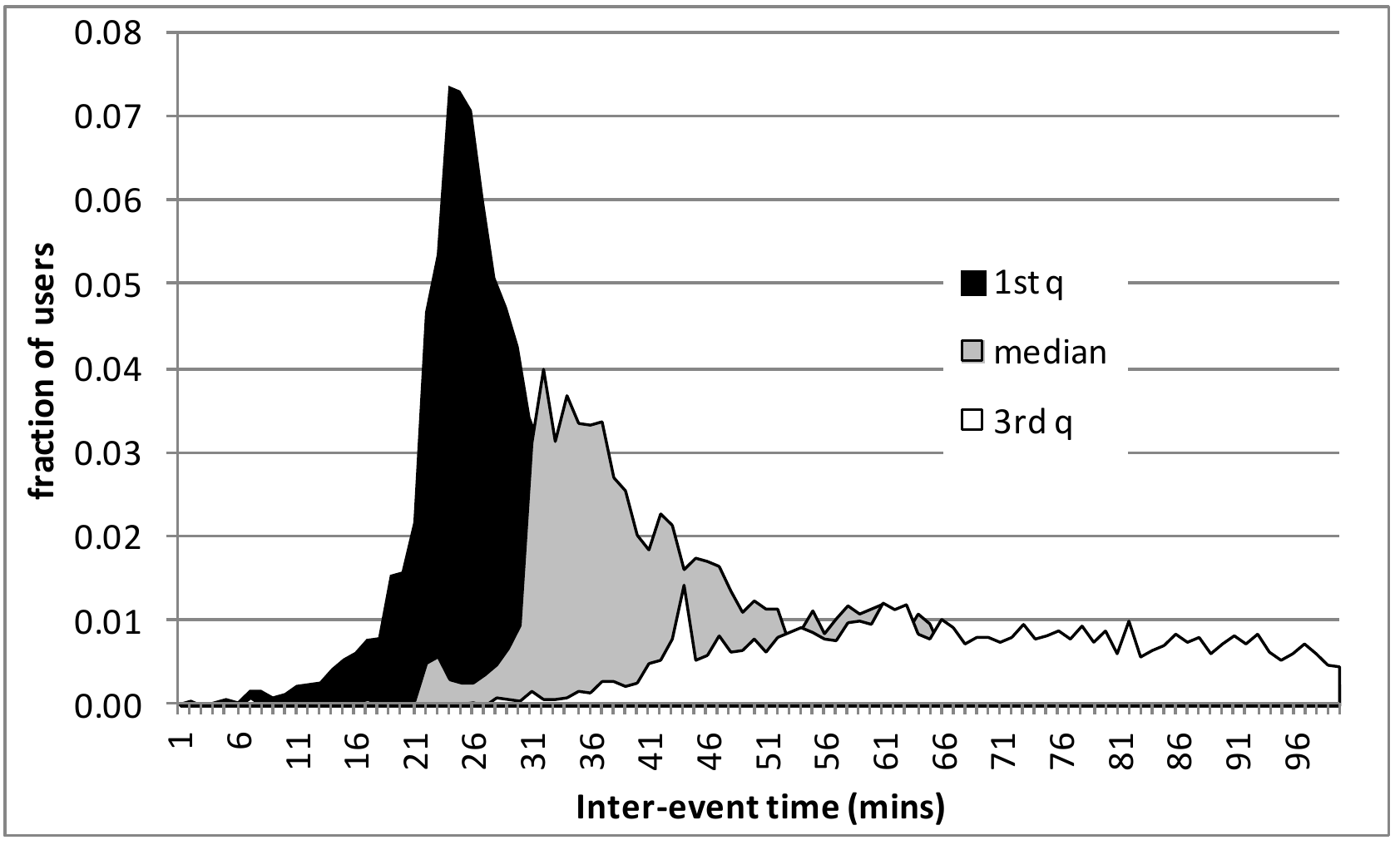}
\end{center}
\caption{Characterization of individual calling activity for the
population possibly attending an event, in terms of time between two
network connections at event time. Graphs show the
distributions of the median (red), first quartile (blue), and third
quartile (green) of individual inter-CDR time}
\label{fig:interevent-time}
\end{figure}

Because the distribution of inter-CDR times for a user spans
several temporal scales, we further characterized each calling
activity distribution during the event time by its first and third
quartile and the median. Figure \ref{fig:interevent-time} shows the
distribution of the first and third quartile and the median for all
the possible attendees. The arithmetic average of the medians is 64
minutes (the geometric average of the medians is 51 minutes) with
results small enough to detect changes of location where the user
stops for about 2 hours.

Such a time frame should be compatible with the duration of a lot of
the events of interest. We also verify that the above figures are
consistent considering also other events.

On the basis of this analysis, we developed the
following approach. We extract CDRs of all the possible
attendees to an event, i.e., all the users that generate at least a
CDR in the event area at the event time. Then, for each user:

\begin{enumerate}

\item We compute the user's average inter-CDR time $iet$ in the
daily hours in which the event takes place. We also compute the time
of the $first$ and of the $last$ CDRs produced in the event area
during the event time.

\item We compute the fraction of time in which the user
is at the event. as:

$f1 = \frac{|last - first + iet|}{event duration}$

\item We then compute, in the same way as before, the fraction of time in which the user is in the
event area in a period spanning $d$ days before the event (in our
experiments we usually set $d=6$). In particular, we compute the time
of the $first$ and of the $last$ CDRs produced in the event area in the $d$ days.

$f2 = \frac{|last - first + iet|}{d \cdot days}$

This represent the fraction of time in which the user is in the
event area without the event.

\item We compute the probability of the user being at the event as

$p = f1 \cdot (1-f2)$

For example, if the user was at the event for the whole event
duration and (s)he never visited that area otherwise, then $p=1$.
Viceversa, if the user is always in the event area $p=0$ because
(s)he is likely to be there for other reasons than the event.

\end{enumerate}

We then add all such probabilities $p$ together to obtain a raw
attendance estimator of the event. It is worth noticing that, in
contrast with other approaches, we do not set a threshold to decide if
a user was present or not. By adding the users' probabilities, it might
happen that 2 users who attend the event with 50\% probability are
considered as 1 user attending with 100\% probability. See Algorithm \ref{algo:radius-att-est}.

\begin{algorithm}[h]
 \KwData{$cdr[\;],bestR,ec,st,et,d=6$}
 \KwResult{$attendance$}
 $candidates[] = usersIn(ec,bestR,st,et)$

 \ForAll{$c_i \in candidates$}{
    $iet = avg-inter-CDR-time(c_i,cdr)$
    $first = timeFirstCDR(ec,bestR,st,et)$
    $last = timeLastCDR(ec,bestR,st,et)$
    $f1 = \frac{|last - first + iet|}{event duration}$

    $first = timeFirstCDR(ec,bestR,st - d_{days},et)$
    $last = timeLastCDR(ec,bestR,st - d_{days},et)$
    $f2 = \frac{|last - first + iet|}{d_{days}}$

    $p_i = f1 \cdot (1-f2)$
 }
 $attendance = \sum_i p_i$
 \caption{Attendance estimator}
 \label{algo:radius-att-est}
\end{algorithm}

\subsection{Linear Regression}

The above estimator is typically much lower than the actual
attendance. This can be naturally explained by the fact that not all
the users will use the phone during the event, and by the fact that
not all of them adopts the same carrier providing the data for this
analysis. In any case, as we will show in the next section, it has a
strong {\it positive} correlation with groundtruth head-counts.
Accordingly, a simple linear regression can scale up the above count
to the actual attendees estimate.

Rather than more complex regression algorithms, we applied linear
regression for two main reasons:

\begin{enumerate}

\item The goal of this work is to show that events' attendance can
be {\it measured} by CDRs coming from the cellular network. If
this is true, then an estimator based on CDR needs only to be
scaled up to provide good results. More complex regression
algorithms could hide shortcomings of the CDR estimator that we want
instead to analyze.

\item The number of events for which we have groundtruth information
is limited.
Accordingly there is a notable risk of overfitting. Regression
mechanisms more complex than linear regression would be even more
susceptible to this problem.

\end{enumerate}

More in detail, we assume the availability of a {\it training} set of events to be used to fit the parameters of the linear regression.
The resulting coefficients are then used to scale CDR estimates of attendance in a {\it testing} set of events.
The combination of all the above steps produces the final estimate of the number of attendees. In the next section, we conduce some experiments to assess the performance of our approach.

\section{Analysis and Results}

As already introduced, to test the performance of the presented
methodology we try to estimate the number of attendees to several events ranging from football matches in stadiums to concerts and festivals in open squares. The analysis spans large events with ground truth attendance of more than 80000 persons to smaller ones
with a ground truth attendance of less than 2000 persons. Overall, we considered a dataset comprising 43 events.

To take into account the fact that a number of CDRs might
happen before and after the event, we set the event starting-time
two hours before the official kick-off, and the event end-time two
hours after the end of the event.

\subsection{Best Radius}

\begin{figure}
\begin{center}
\includegraphics[width=1.0\columnwidth]{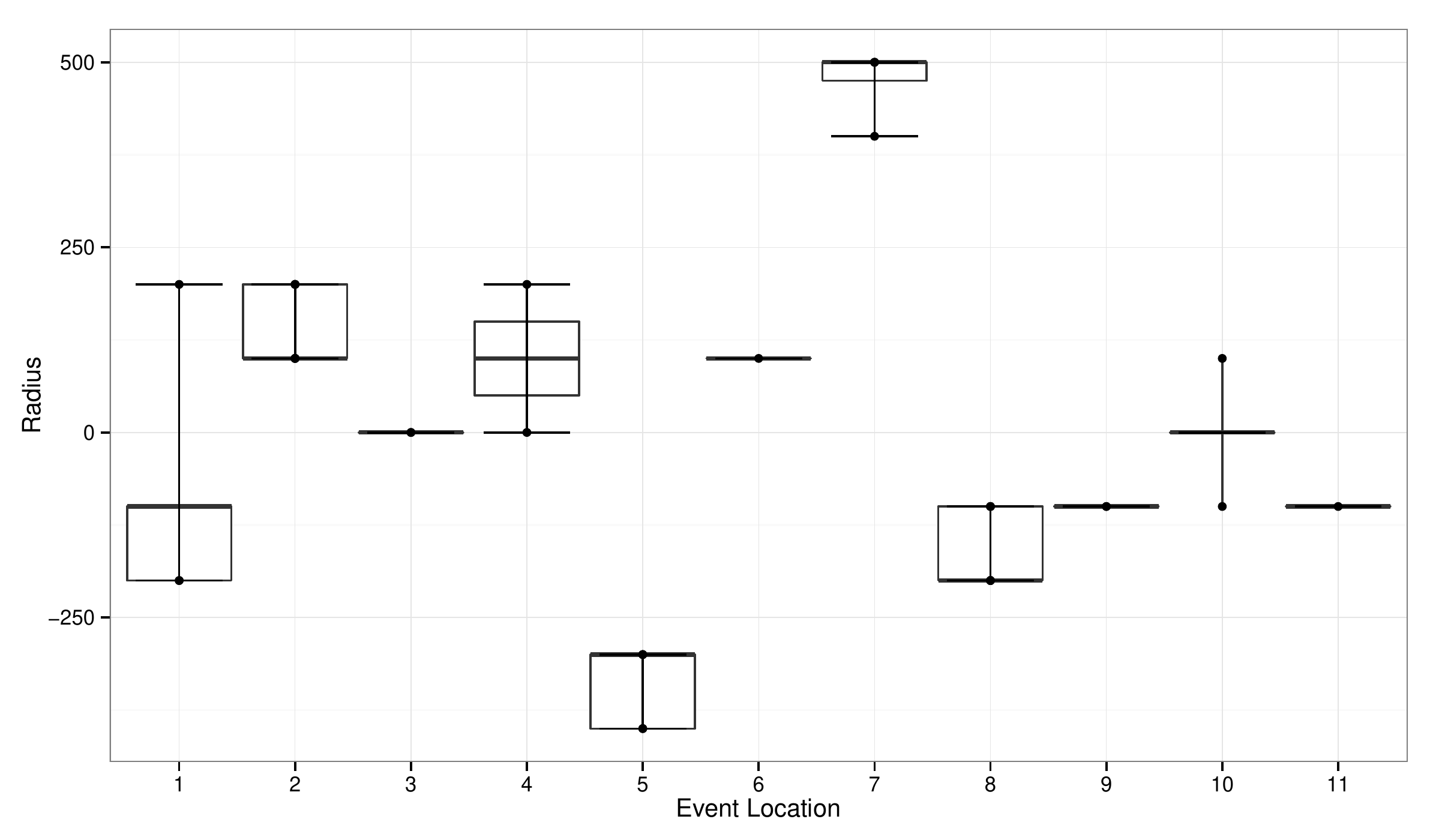}
\end{center}
\caption{Diagram showing best radius results for different places. It is possible to see that different events in the same place can be represented by different radii}
\label{fig:best-radii}
\end{figure}

In this first set of experiments we report the radius that best capture the dynamics of a given event.
We run the algorithm described in Section 3.2. Specifically, we varied $r$ in $r_k \in
[-500m,1500m]$ with a $100m$ step. The result is a set of $NR = 21$ radii
to be tested. Figure \ref{fig:best-radii} illustrates the obtained results for different event areas under analysis (on the x-axis we indicate an id associated to different event areas -- e.g., 1 = ``a stadium in Bergamo, Italy'') It is possible to see that the same event area may be best represented with  different radii depending on the specific event considered. This is rather natural, as the more people attend the event, the more the cells nearby the stadium get saturated and rely the network connection to farther cells. Accordingly, larger events (even in the same location) tend to be associated to larger radii.

\subsection{Attendance Estimate}

In this set of experiments we actually estimate attendance for multiple events. First we use the algorithm described in Section 3.3 to obtain a CDR count proportional to the attendance estimate. Then we scale that number with a linear regression. Specifically, for each event to be analyzed, we considered as a training set all the events happening in stadiums (leaving out the considered event, if present). We use the estimated attendance of such events and the associated groundtruth attendance information to fit the parameters of a linear regression. We use events in stadiums as training set as they are typically associated with better groundtruth estimates (derived from ticketing information). We then scale the CDR count with the linear regression to obtain the final estimate. Specifically, we report results using different kinds of linear regression:

\begin{enumerate}

\item {\bf Standard linear regression.} In this approach, we consider the whole training set, create a linear regression model fitted by minimizing sum of squared errors, and use the model parameters to scale predicted attendance count.

\item {\bf Piecewise linear regression.} In this approach, for each testing sample, we consider the $n$ closest samples in the training set, create a linear regression on that $n$ points, and use it to scale that predicted testing sample. In our experiments we empirically set $n = 6$.

\item {\bf Range linear regression.} We also conducted some experiments separating the events with an attendance below and above 10000 persons. This can be interpreted as a trade-off between global and piecewise regressions: we fit one regression for small ($< 10000$ persons) events, and another for large events ($\geq 10000$ persons).

\end{enumerate}

\begin{figure*}[!htb]
\begin{center}
\includegraphics[width=0.66\columnwidth]{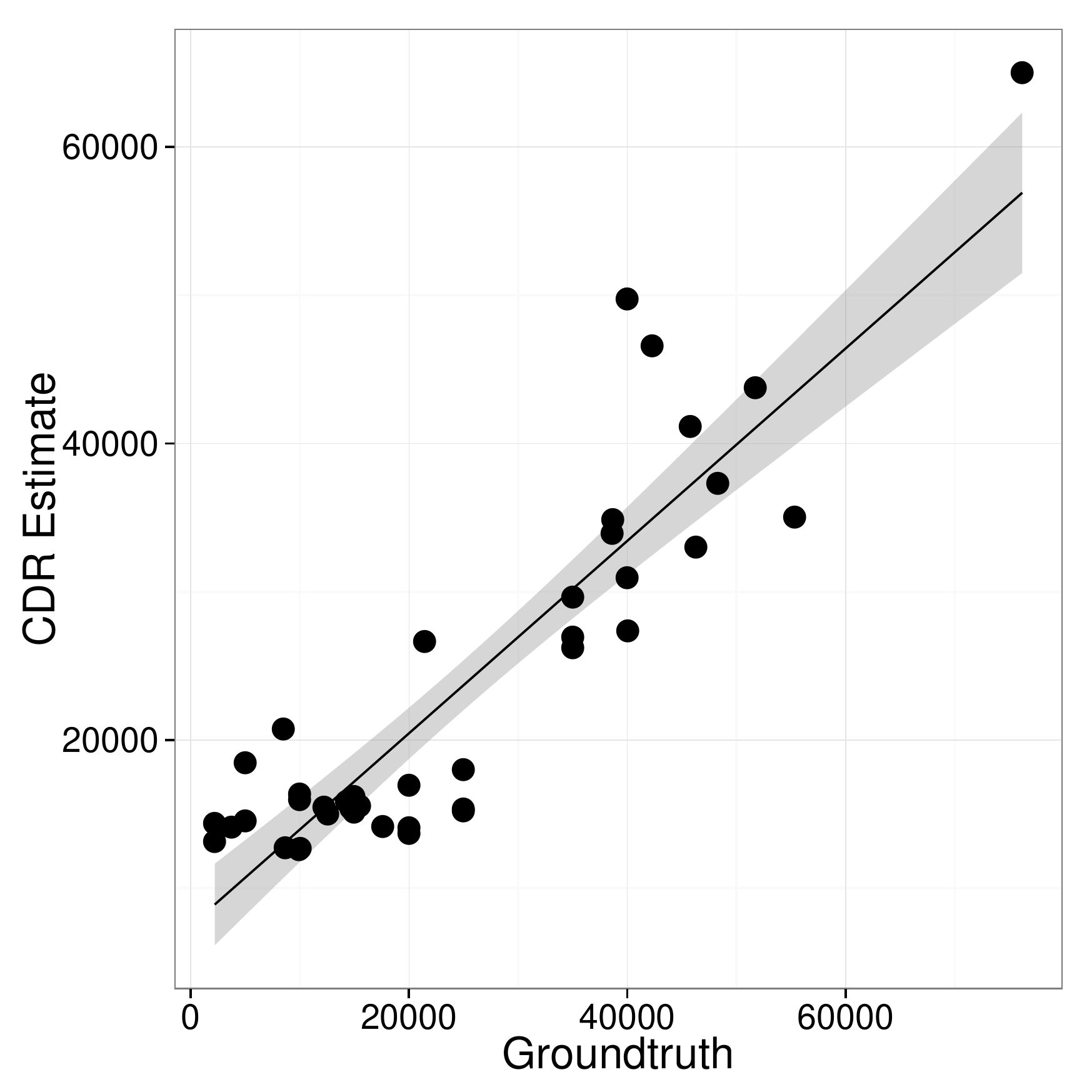}
\includegraphics[width=0.66\columnwidth]{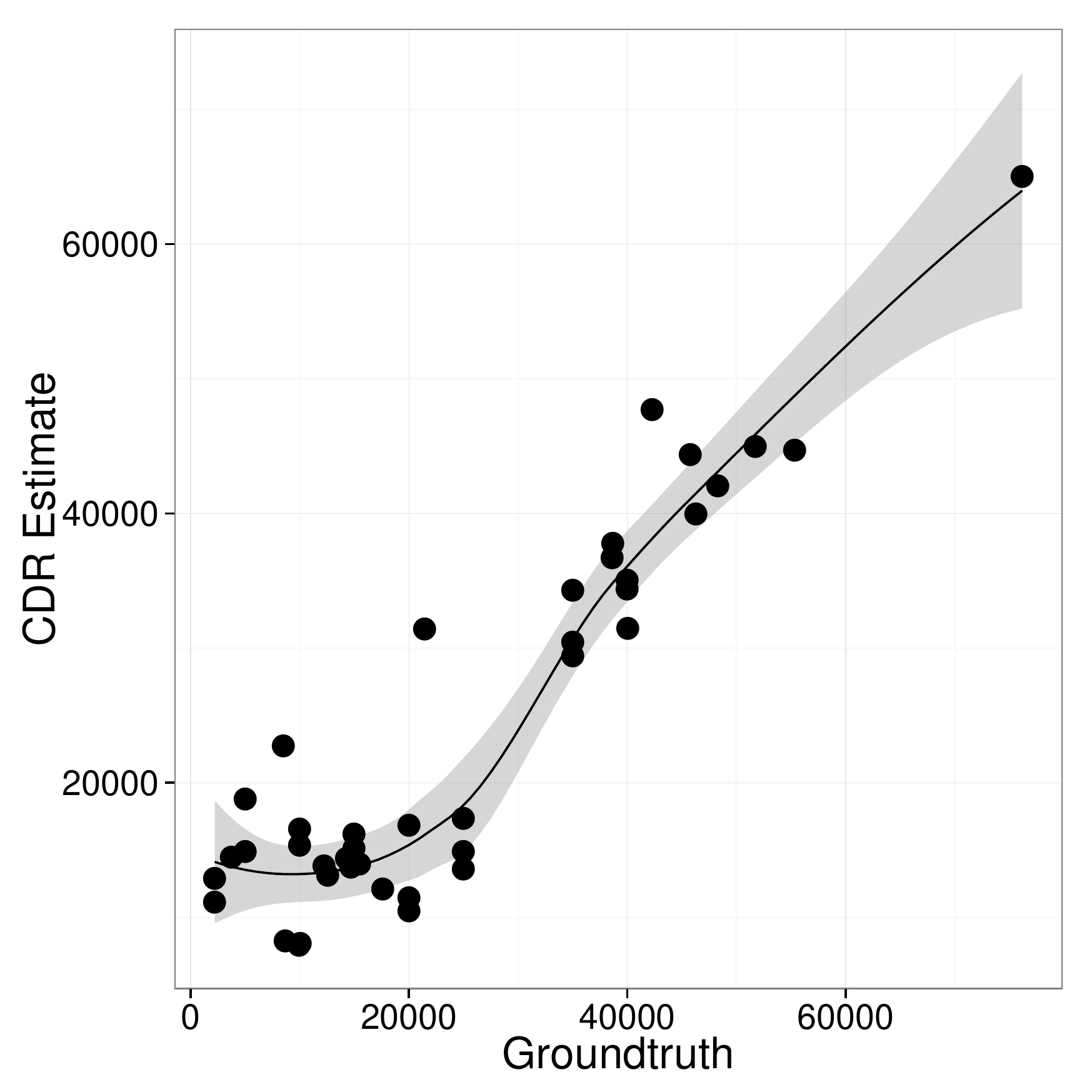}
\includegraphics[width=0.66\columnwidth]{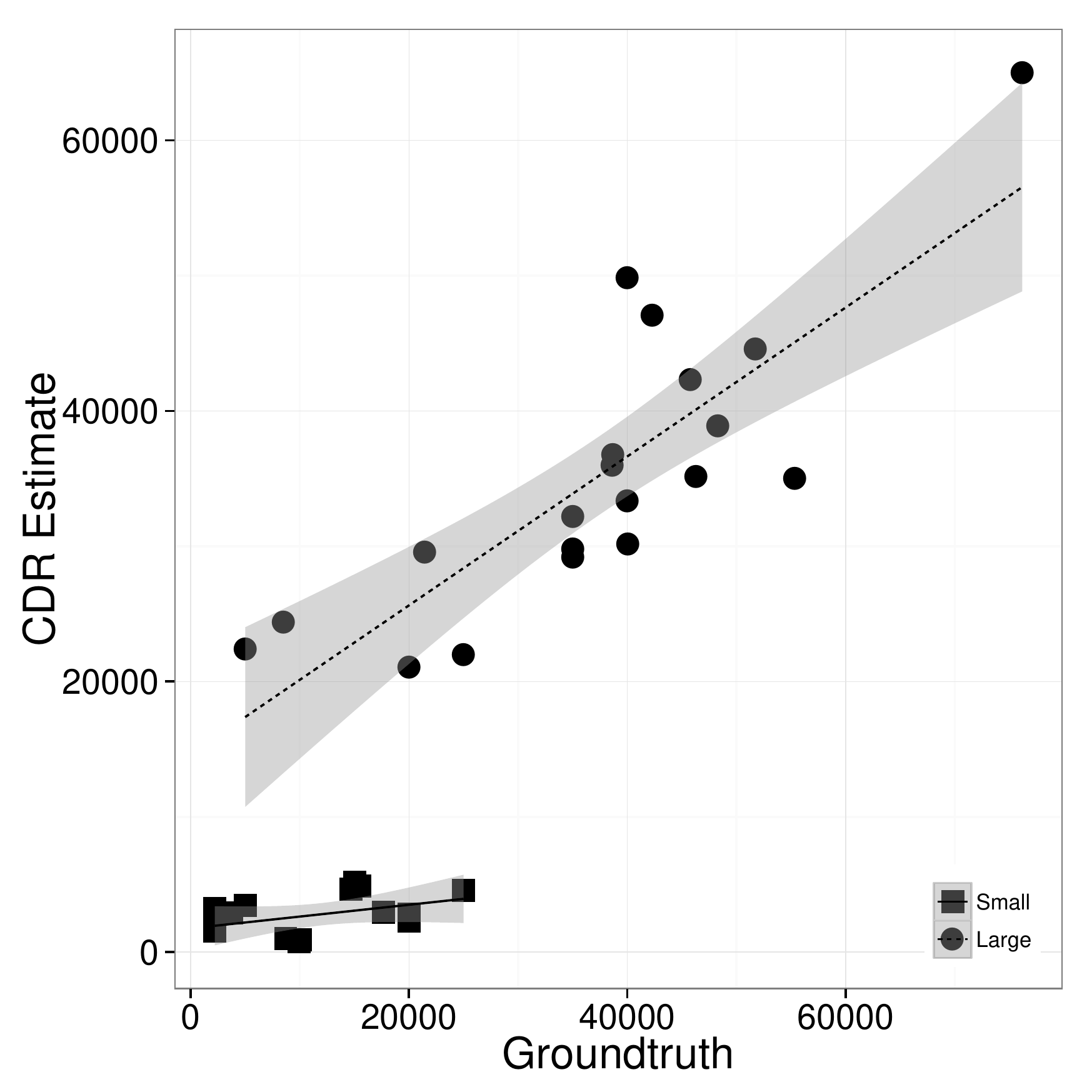}
\end{center}

\setlength\parindent{42pt}
\parbox{0.66\columnwidth}{

\begin{tabular}{| l | r |}
  \hline
  {\bf Linear R.} & $r^2=0.76$\\
  \hline
  Mean abs. err. & 6378 \\
  Median abs. err. & 5447 \\
  Mean \% err. & 73\% \\
  Median \% err. & 25\% \\
  \hline
\end{tabular}
}
\parbox{0.66\columnwidth}{
\begin{tabular}{| l | r |}
  \hline
  {\bf Piecewise R.} & \\
  \hline
  Mean abs. err. & 6057 \\
  Median abs. err. & 4072 \\
  Mean \% err. & 68\% \\
  Median \% err. & 14\% \\
  \hline
\end{tabular}
}
\parbox{0.66\columnwidth}{
\begin{tabular}{| l | r |}
  \hline
  {\bf Range R.} & \\
  \hline
  Small Events & $r^2=0.42$\\
  \hline
  Mean abs. err. & 5024 \\
  Median abs. err. & 3455 \\
  Mean \% err. & 66\% \\
  Median \% err. & 68\% \\
   \hline
  Large Events & $r^2=0.86$\\
  \hline
  Mean abs. err. & 6032 \\
  Median abs. err. & 4841 \\
  Mean \% err. & 39\% \\
  Median \% err. & 14\% \\
  \hline
\end{tabular}
}
\caption{Attendance estimation results. ({\bf top row}) correlation plot with different kinds of regressions. (left) linear, (center) piecewise, (right) range. The shaded area represents confidence interval for the regression outcome. ({\bf bottom row}) $r^2$, mean/media absolute and percentage errors for the different regressions.} \label{fig:results}
\end{figure*}

Figure \ref{fig:results}(top-row) illustrates
the result of the different regressions between groundtruth data and our
attendance estimator. Other than visually, we verified that in the case of linear regression (left plot), the
results exhibit a Pearson correlation $r=0.87$ and a coefficient of
determination $r^2=0.76$ indicating a strong positive correlation
between the results. In the case of piecewise regression (center plot) summarizing a single correlation coefficient is problematic. However, it is possible to see a good fit of the data. In the case of range linear regression (right plot), $r=0.65$/$r^2=0.42$ for small events ($< 10000$ persons), $r=0.93$/$r^2=0.86$ for large events ($\geq 10000$ persons), indicating offering weak results for small events, while strong correlation for large ones.

In all the plots, confidence intervals for the regression is depicted with a gray area.

Figure \ref{fig:results}(bottom-row) illustrates mean/median absolute error between estimated attendees and groundtruth, and mean/median percentage error (absolute error divided by groundtruth).
The gap in errors between mean and median indicates that the distribution of error is skewed (in the case of linear regression, skewness = 3.10, in the case of piecewise linear regression skewness = 2.69, in the case of range regression, skewness = -0.6/3.3 for small and large events respectively).  This is due to the fact that even small errors in the order of 1000 person would be very high in events with 2000 attendees (50\% error) thus notably increasing the mean error.

\begin{figure*}[!htb]
\begin{center}
\includegraphics[width=0.66\columnwidth]{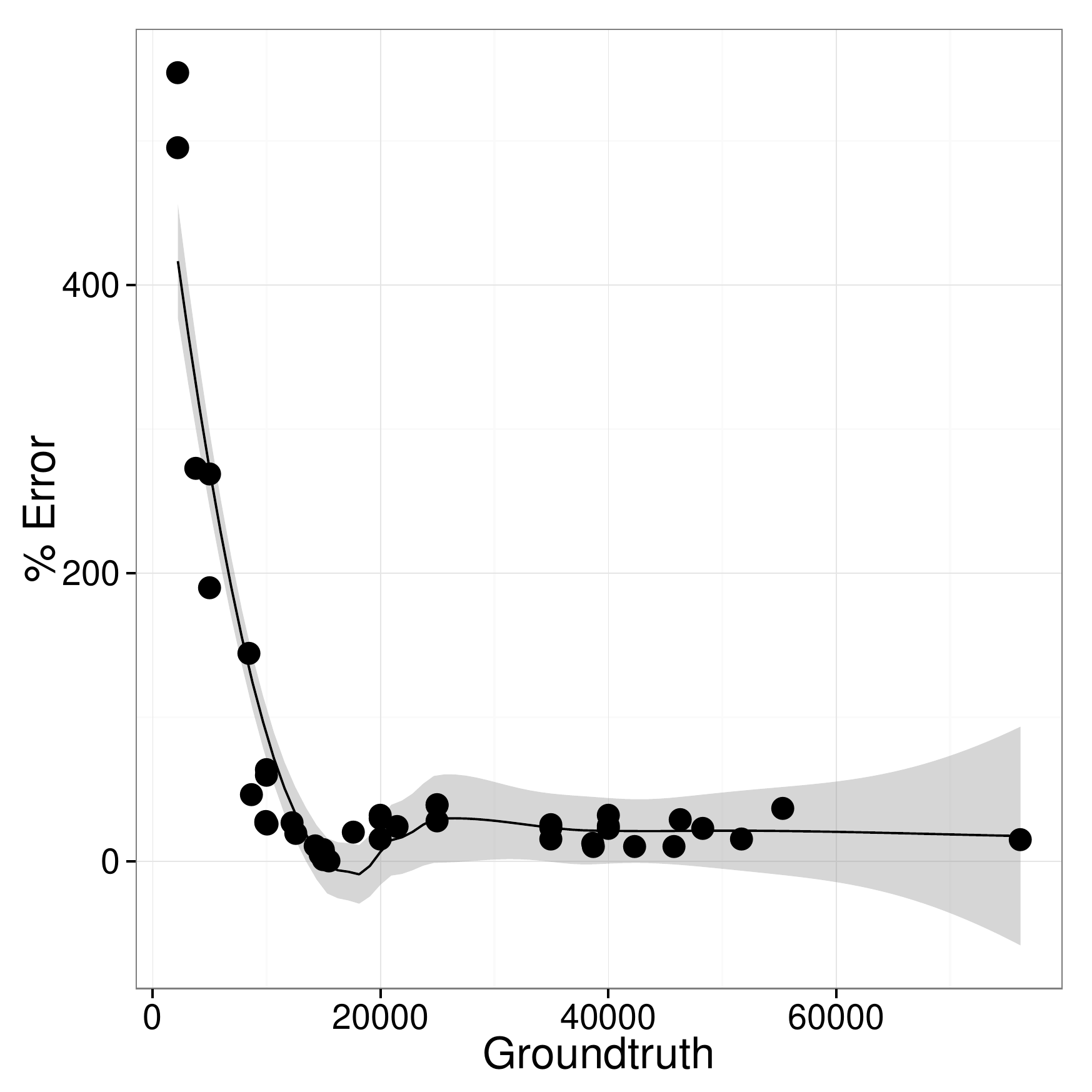}
\includegraphics[width=0.66\columnwidth]{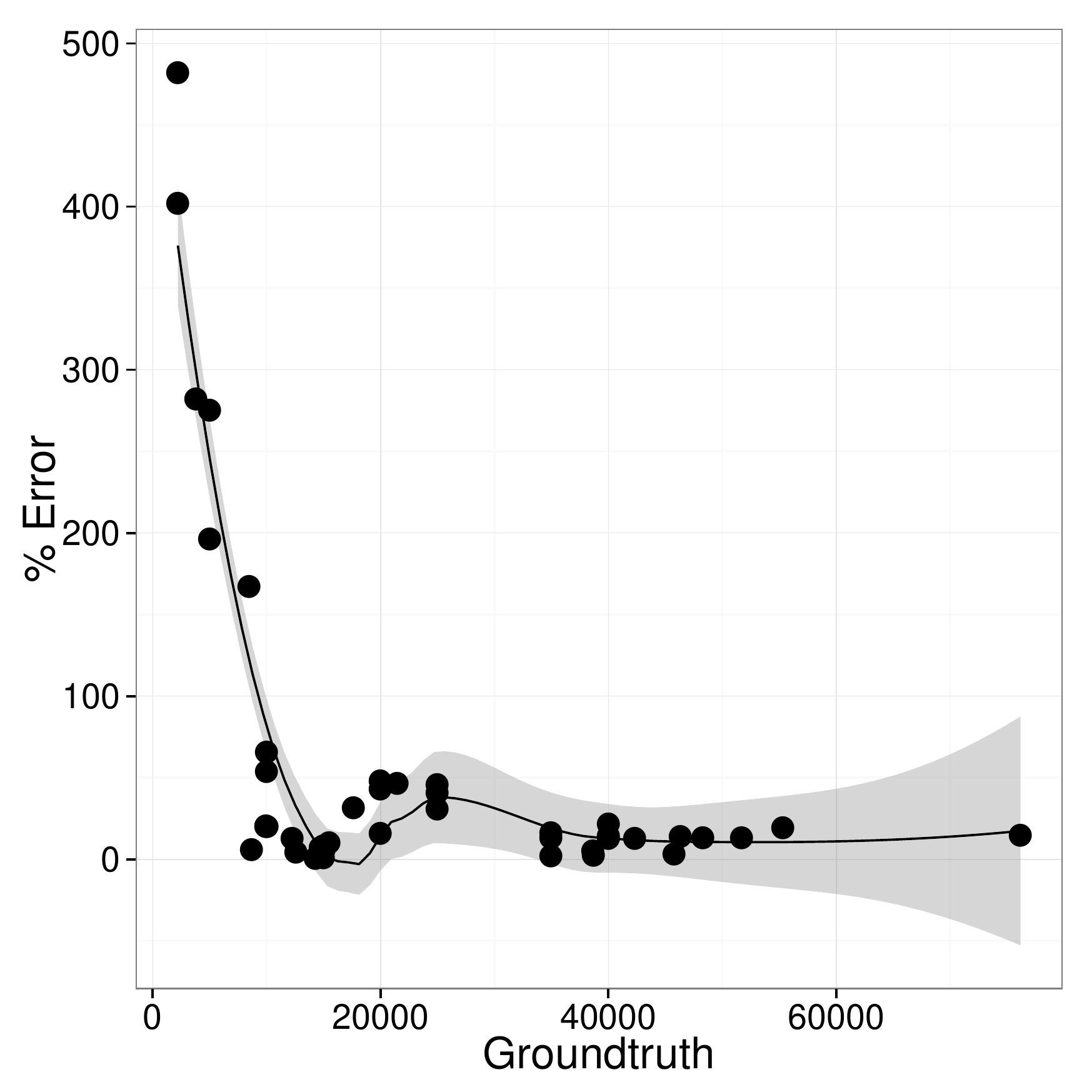}
\includegraphics[width=0.66\columnwidth]{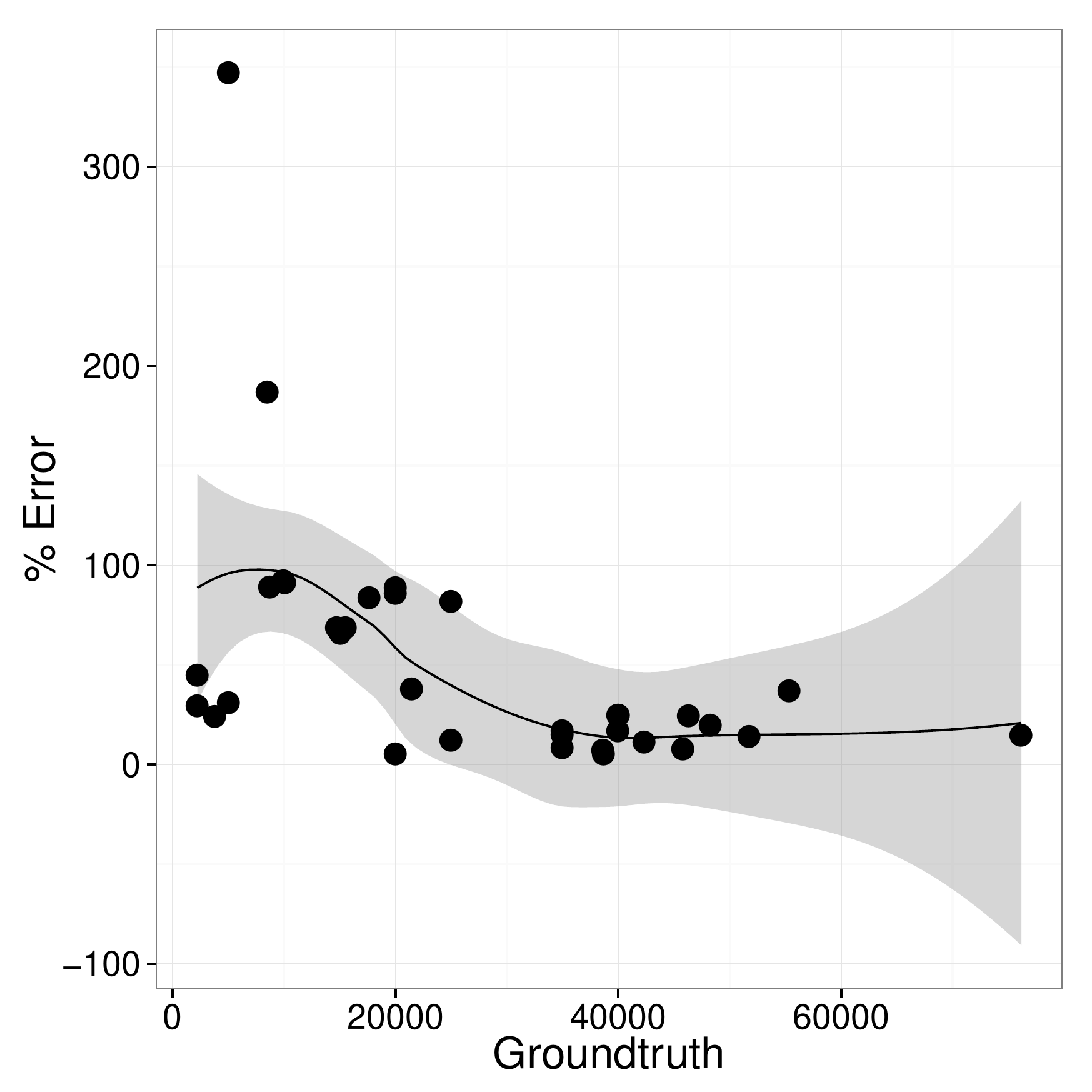}\\
\includegraphics[width=0.66\columnwidth]{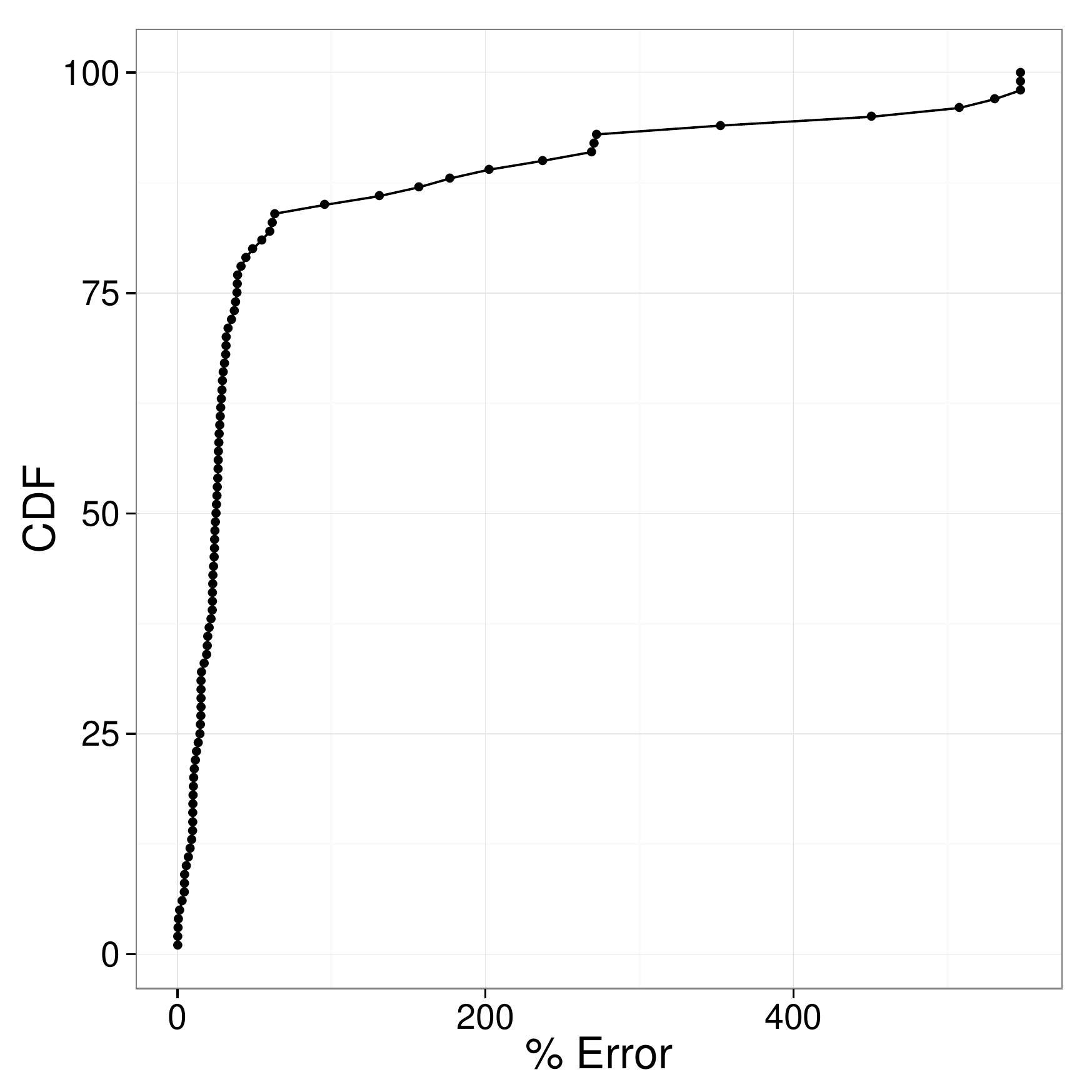}
\includegraphics[width=0.66\columnwidth]{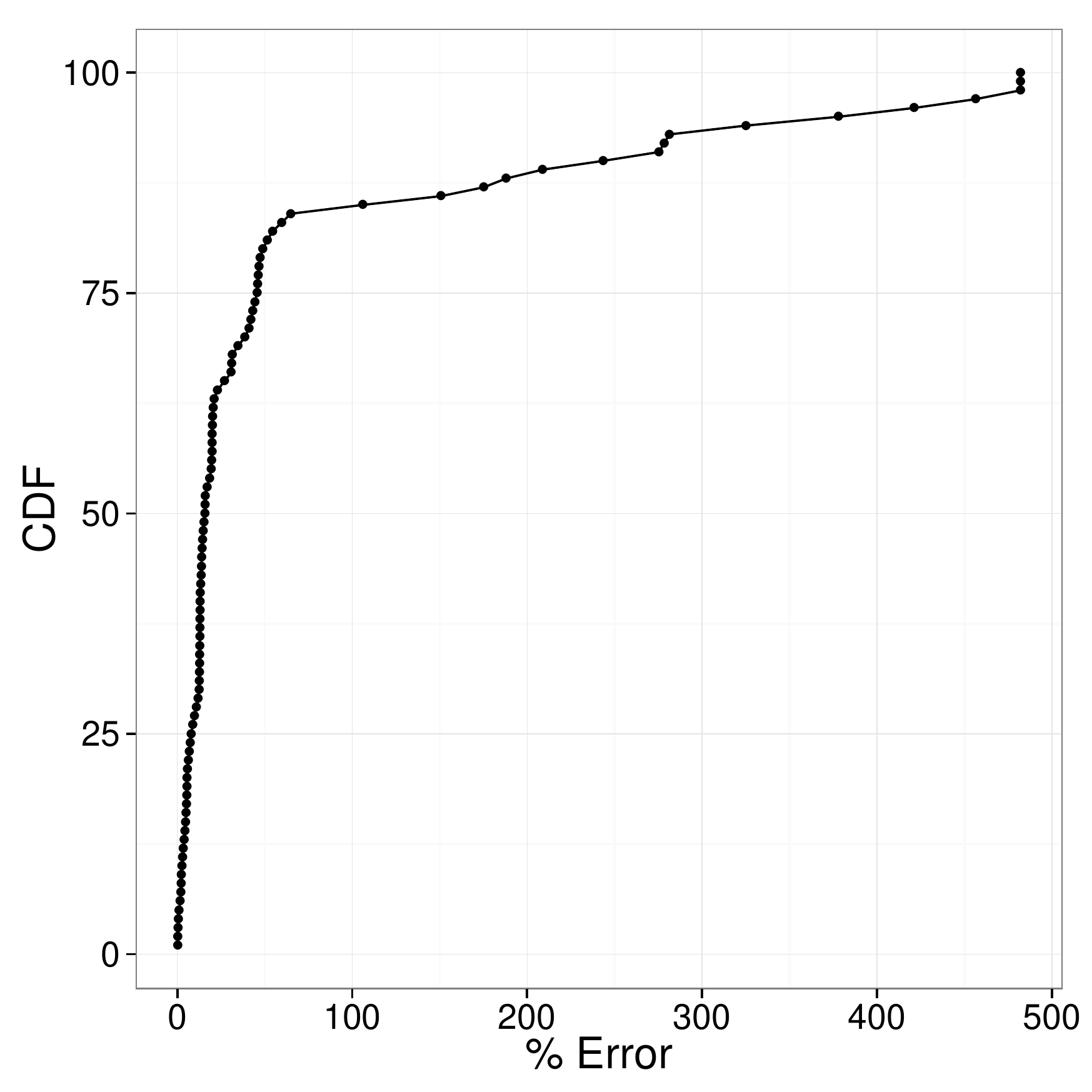}
\includegraphics[width=0.66\columnwidth]{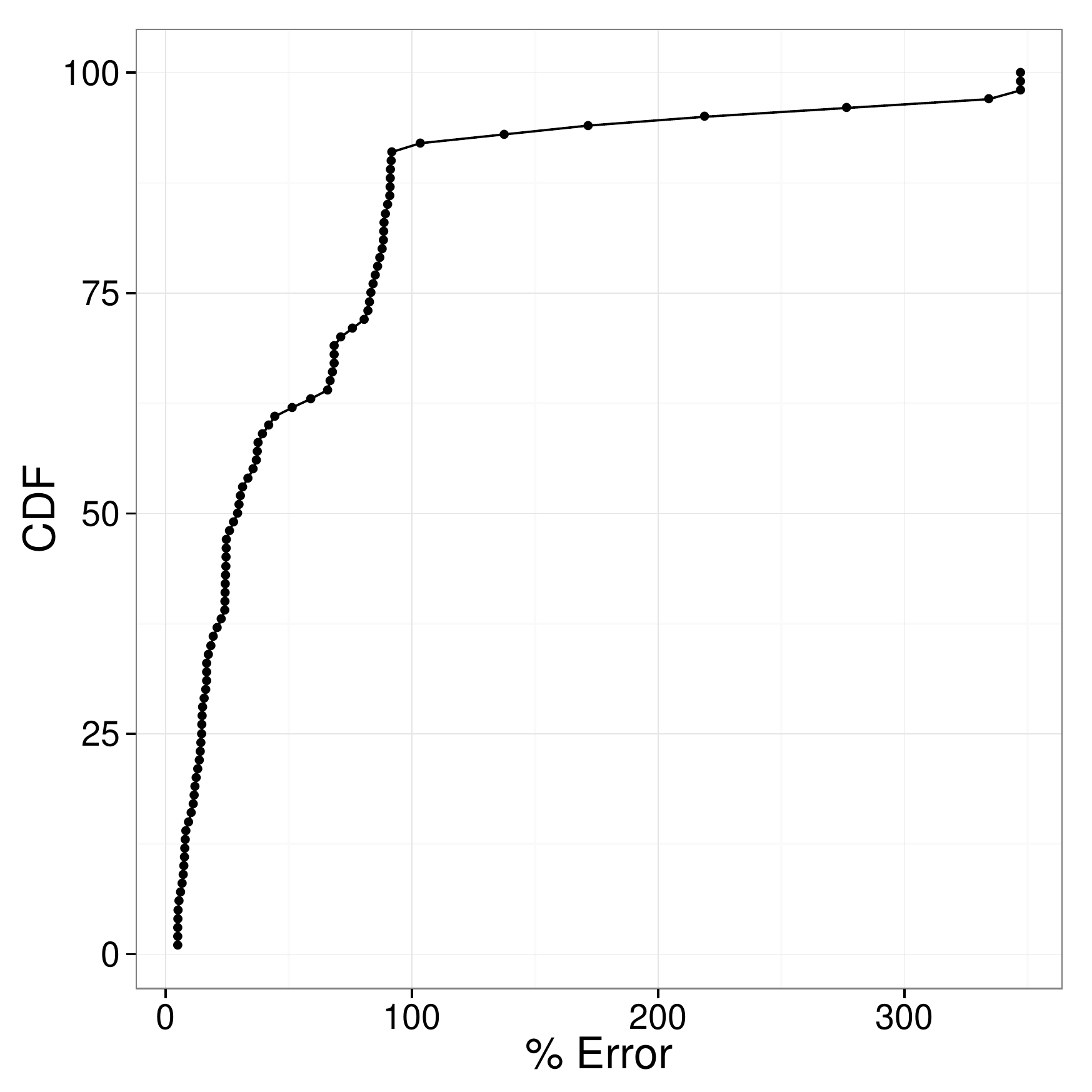}
\end{center}
\caption{Attendance estimation errors. ({\bf top row}) error distribution for different regressions: (left) linear, (center) piecewise, (right) range. ({\bf bottom row}) error cumulative density function. It is possible to see that error distribution is highly skewed. Our approach is effective for large events, while it has considerable errors for small ones.} \label{fig:error-evaluation}
\end{figure*}

To better quantify this behavior, Figure \ref{fig:error-evaluation} shows error distribution with regard to groundtruth attendance (top-row) and the error CDF (bottom-row). The graph shows results for linear regression (left), piecewise linear regression (center), range regression (right). Looking at the graph, it is easy to see the skewness effect described above. In all the regressions, rather expectedly, the approach presents large errors for small events, while small errors for large events.

In summary, it is possible to see that the use of the described approach produces rather good estimates of the number of attendees. It is easy to see that results are better in large events where a limited absolute error has a small impact in the overall percent error. In general, we found that the proposed approach starts producing consistent good results for events larger than 10000 attending persons.
Considering only those events with an attendance greater than 10000, Pearson correlation jumps to 0.93. Linear regression's  mean \%error drops to 22\% and median \%error drops to 15\%. Similarly, piecewise linear regression's  mean \%error drops to 16\% and median \%error drops to 13\%.

\subsection{Unstructured Events}

The dataset of events used for the experiments comprises two kinds of events: {\it (i)} ``structured'' events, like concerts and football matches, for which some sort of entrance policy (e.g., entrance gates) allow to obtain reliable estimate on the number of attending persons. {\it (ii)} ``unstructured'' events happening in open squares or parks for which no entrance policy is enforced. The analysis of this latter kind of events is problematic because it is very difficult to obtain reliable groundtruth attendance estimates, however -- for the same reason -- it is also the best scenario for the actual use of the proposed technique.

\begin{figure}
\begin{center}
\includegraphics[width=0.66\columnwidth]{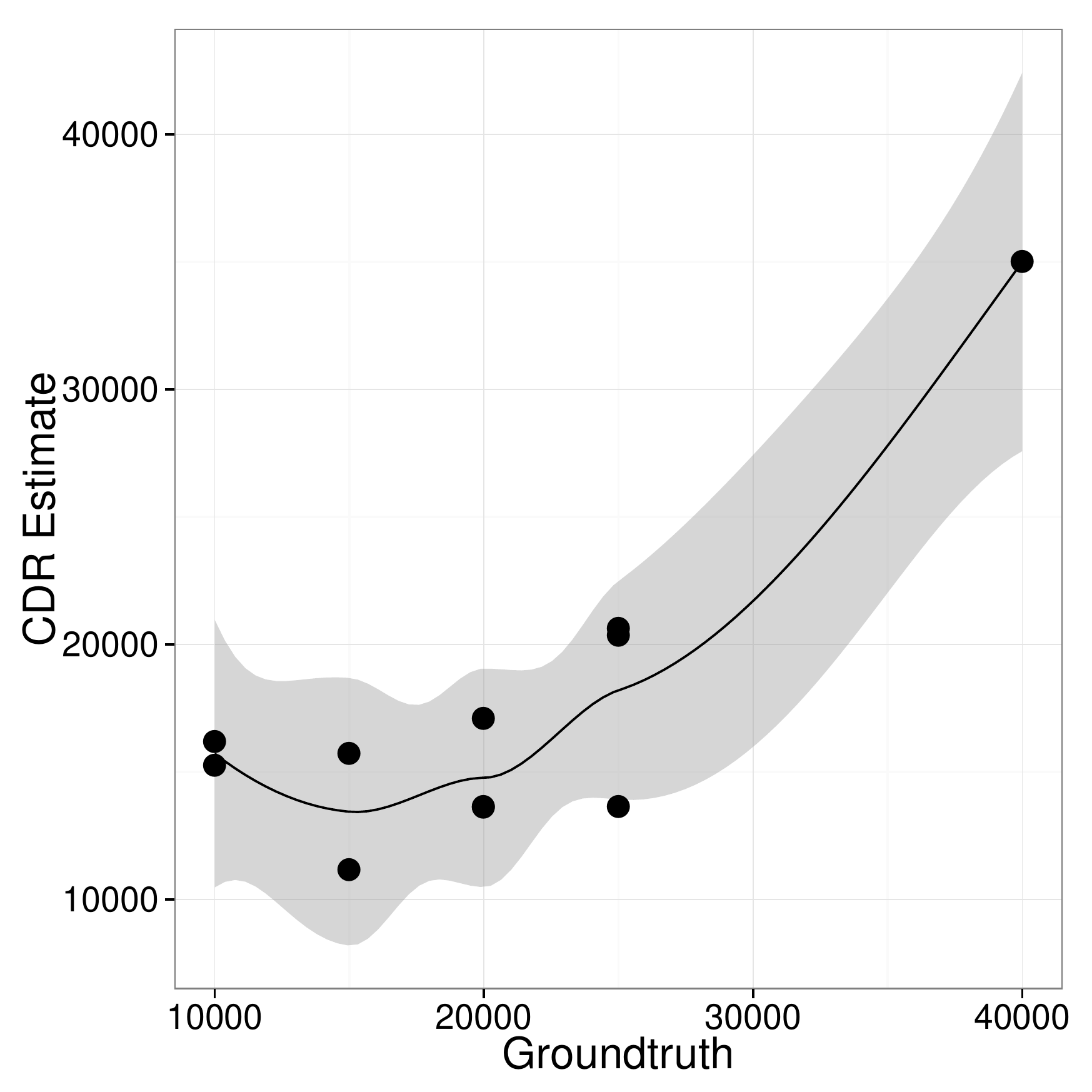}
\end{center}

\setlength\parindent{70pt}
\begin{tabular}{| l | r |}
  \hline
  {\bf Piecewise R.} & \\
  \hline
  Mean abs. error & 7307 \\
  Median abs. error & 4422 \\
  Mean \% error & 33\% \\
  Median \% error & 22\% \\
  \hline
\end{tabular}

\caption{Results obtained by using piecewise linear regression for ``unstructured'' events} \label{tab:unstructured-results}
\end{figure}

Figure \ref{tab:unstructured-results} illustrate results for a set of ``unstructured'' events. We fit the linear regression by using ``structured'' events (football matches in stadiums) happening in the  same city and we searched the Web for reported attendance estimates and use them as ground truth. In this case results are worse than in the previous case, obtaining 22\% median \%error. On the one hand, this is due to the fact that these events tends to be smaller than ``structured'' ones, thus making the attendance estimation task inherently more difficult. On the other hand, the linear regression is trained for larger ``structured'' events, thus it can be a less effective fit for these events. Finally, as groundtruth estimate for this events is weaker, the fact that a number of events have an estimated attendance {\it lower} than the groundtruth might be also interpreted as the fact that the estimates reported in the news (on the Web) are inflated.


\section{Knowledge of Multiple Events}
\label{sec:multiple}

In all the previous algorithms and experiments we considered events in isolation: we tried to estimate the attendance to an event {\it without} any information about other events happening in the same place.
On the contrary, if we know a number of events that happened in a given place, we can adopt a different procedure to estimate the radius of the event area.

In particular, we try to estimate the area associated to a given placemark (e.g., a stadium), and all the events happening in there will be associated to the same event area.
This procedure updates the procedure described in Section \ref{sec:radius} STEP 2. The idea is that, instead of weighing each possible radius by how extreme values (z-score) it produces, we weight each radius by how many events it is able to identify as outliers. This is basically the procedure in Figure 6 in which we count the number of events. More in detail the process to identify the event radius is the following:\\

\noindent
{\bf STEP 1.} The same as in Section \ref{sec:radius}\\

\noindent
{\bf STEP 2.}

\begin{enumerate}

\item For each event area, we identify a number of events happened in there (this will serve as a sort of ``training" set).

\item We consider the {\it z-score} computed in STEP 1 for a time frame encompassing all the events in the training set.

\item We identify outliers in the $z_i$ values as those points with a value greater than
$3$. We then count the number of outliers that happens to be at the
same day and time of events in the ``training'' set. The result is
that for each value $r_k$ we have the number of events being
identified $e_k$.

\item The final value of the radius $best\_r$ is the average of the
$r_k$ values weighted by the number of identified events $e_k$

\end{enumerate}

\begin{algorithm}
 \KwData{$z[\;]$}
 \KwResult{$bestR$}
 \ForAll{$r_k \in [r_{min}, r_{max}]$ }{
    $x_k = countUsers(cdr[\;],ec,r_k,st,et)$

    $e_k =  number\;of\;event\;days\;with\;z_k > 3$

    $bestR = \frac{\sum_k r_k \cdot e_k}{\sum_k e_k}$
 }
 \caption{Multiple events. Radius Extraction - Step 2}
 \label{algo:radius-step2multi}
\end{algorithm}

\begin{figure*}[t]
\begin{center}
\includegraphics[width=0.66\columnwidth]{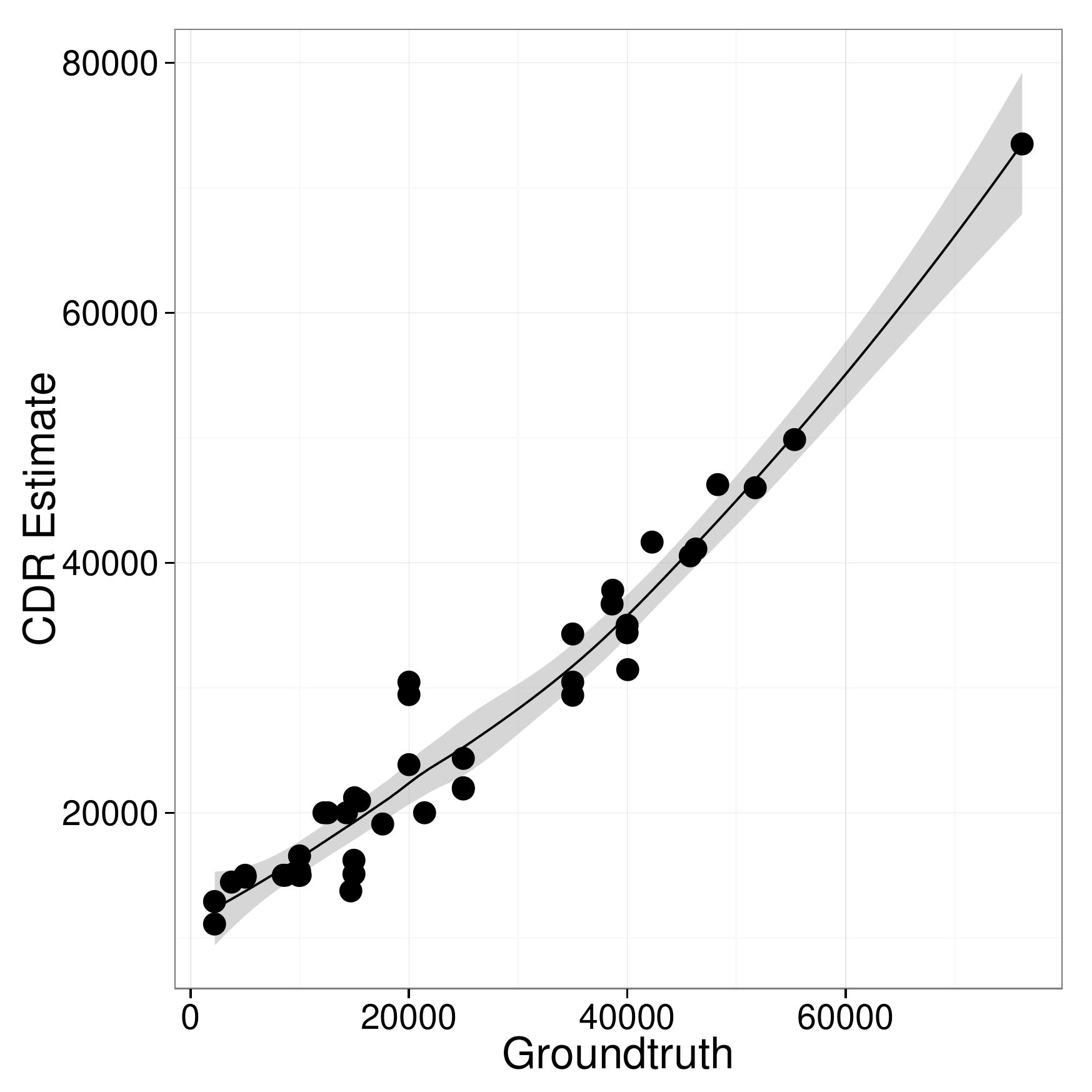}
\includegraphics[width=0.66\columnwidth]{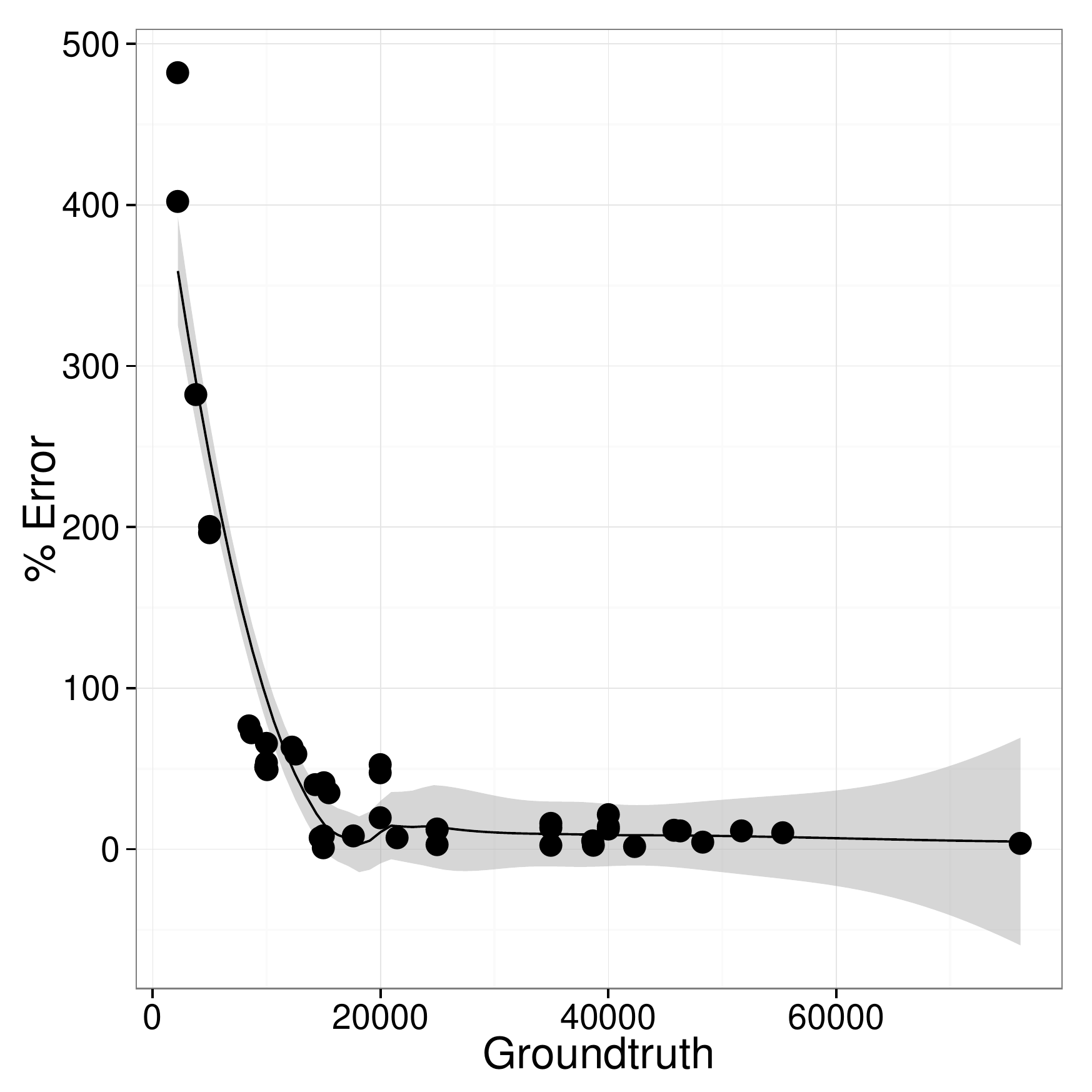}
\includegraphics[width=0.66\columnwidth]{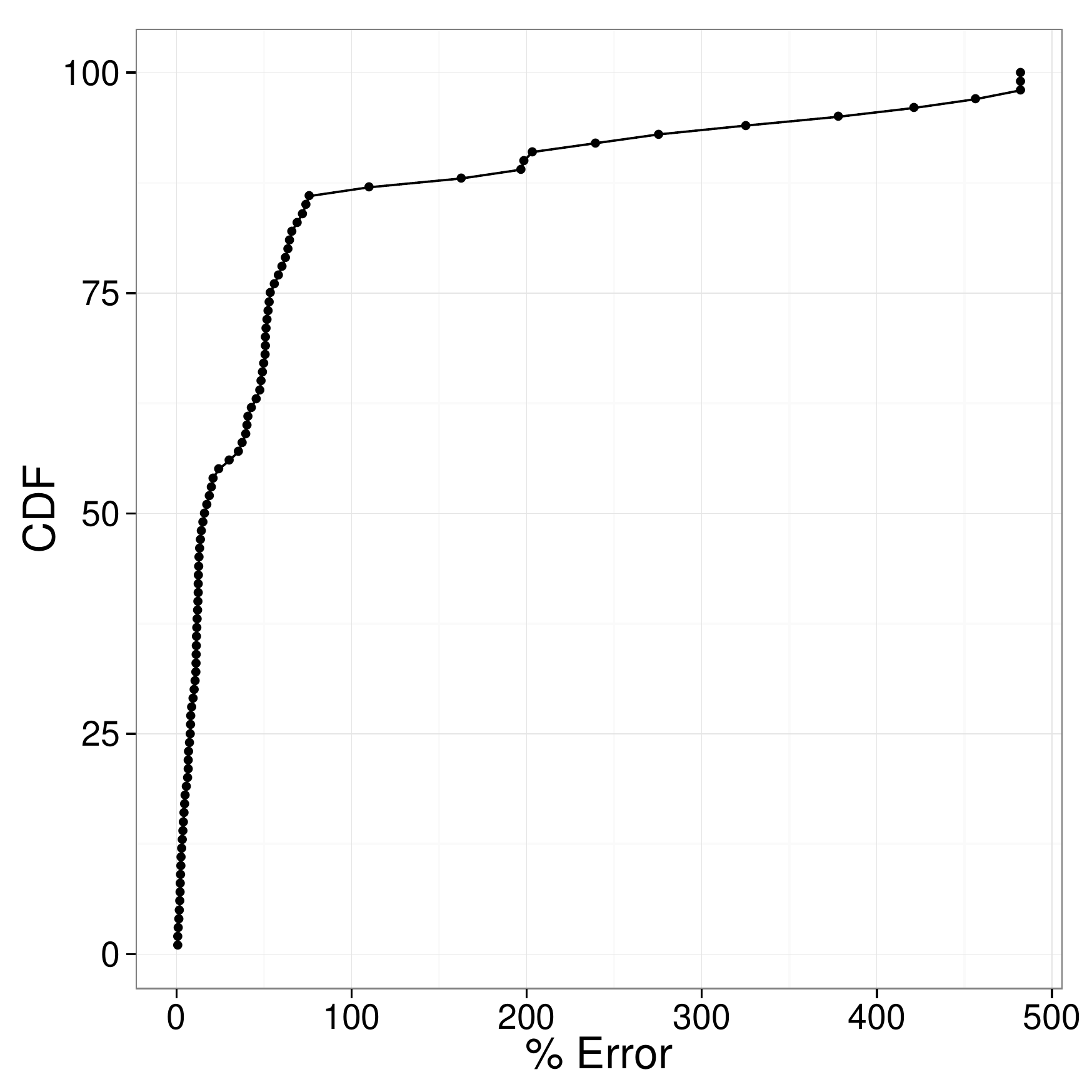}

\end{center}
\caption{Results obtained by adopting the radii computed with the knowledge of multiple events. (left) correlation plot, (center) \% error distribution, (right) \% error CDF}
\label{fig:multiple}
\end{figure*}

See a more formal description in algorithm \ref{algo:radius-step2multi}. On the one hand this approach tends to be more robust in that radius parameters are choose to detect the largest number of events. On the other hand, it is less flexible in that it associates a single radius to a given place without the flexibility of enlarging the radius for larger events in the same place. Figure \ref{fig:multiple} illustrates the results of the estimation approach with the radii computed in this way and adopting piecewise linear regression. (left) correlation plot, (center) \% error distribution, (right) \% error CDF. Overall, the knowledge of multiple events further improves the results: $r = 0.95, r^2 = 0.9$, median absolute error = 4160, median \% error = 10\%.

\section{Related Work}

The application potential of estimating the number of people present in specific parts of the city at specific times led to the development of a number of approaches and researches to tackle the issue.  Such estimates can in fact provide useful information to the local government and to the event organizers to plan, manage and respond to the event. Also the advertisement industry would get notable information from such data, in that it is possible to measure how
many people were able to see a given advertisement, understand where they come from, their habit, etc.

People counts, surveys, and other traditional methods to identify the presence of visitors and tourists in a city are often expensive and result in limited empirical data. Similarly, the exploitation of land-use (e.g. density of hotels) and census data provides only a static perspective on city dynamics. The lack of data presents particular difficulties given that most cities -- though they may aim at providing advanced services -- have limited human, technical, and financial resources. Today, thanks to the emergence of ubiquitous technologies, new data sources are available

Fueled by the ``recent'' availability of telecoms' CDR data, a
number of researchers try to automatically identifying events happening in the city and
estimating the number of people attending the event.

The works in \cite{Gir09,Neu13} present an approach to estimate the
attractiveness of events happening in the city from the combination
of cellular network activity and other information sources. They try to
estimate the location of cellular network traffic and to use it as a
proxy of the number of people in that area. However, these methods can identify daily trends and outliers, but they
can not estimate the actual number of people.

The work presented in \cite{Que11,Cal10} presents another approach
to analyze people attendance to special events on the basis of CDRs coming from the AirSage ({\it www.airsage .com}) platform. In
this work, they segment users' traces to identify those places where
a user stops. If this place, coincides with the place of the event
and if the duration of the stop is at least the 70\% of the duration
of the event, the user is classified as attending the event. On this
basis they are able to analyze the attendance to specific events.
However, they claim: ``{\it Estimating the actual number of
attendees is still an open problem, considering also that ground
truth data to validate models is sometime absent or very noisy}''
and do not perform quantitative analysis in this direction.

The work in \cite{Tra11} is very interesting and closer to our
approach. They use a Bayesian model to localize the source of CDRs. Then, they compute the probability $p$ of each user to
participate an event as $p = p1 \cdot (1-p2)$. Where $p1$ is the
fraction of time in which the user is in the event area at the event
time. $p2$ is the fraction of time in which the use is in the event
area at other times. Finally, they use an outlier detection
mechanism (based on a z-score) to classify users as participants to
an event. Unfortunately, they use the approach only to identify an
event and not to estimate the actual attendance.

A similar approach to identify events is reported in \cite{Fer14}.
In this work, authors apply an outlier detection mechanism to
aggregated cell network data (i.e., erlang measurements). Events are
associated to overcrowded or suddenly underpopulated areas.

In conclusion of this section, while some works propose approaches
to detect and analyze events happening in the city by using the data
from cellular network, the problem of actually estimating the number
of attendees is largely unexplored. In particular, to the best of
our knowledge, there are not published results of the accuracy of
attendance estimation using CDRs.

\section{Concluding Remarks}

In this work we propose an innovative methodology to estimate the
number of attendees to events happening in the city from cellular
network data. We evaluate our approach in 43 events ranging from football matches in stadiums to concerts and festivals in open squares. Comparing our results with the best groundtruth data available, our estimates provide a median error of less than 15\% of the actual number of attendees.

While the obtained results are very encouraging, there are a number of research directions that could improve the presented work:

\begin{itemize}

\item Of course, running experiments on other, more diverse, events would better validate our results.

\item Our work has been mainly driven by experiments. A better theoretical framework for our modeling (especially with regard to the event area estimation) could provide further ideas for improvement.


\item A deeper analysis of the trajectories of individual users could provide a more fine grained localization of CDRs, thus leading to a better estimate of the user's presence in the event area \cite{Leo14}

\end{itemize}

Despite the above limitations, to the best of our knowledge, this is the first work  providing a practical and accurate way of estimating the number of attendees to events happening in the city from cellular
network data.

\bibliographystyle{elsarticle-num}
\bibliography{d1}

\end{document}